\documentclass{aa}
\usepackage[varg]{txfonts}
\usepackage{graphicx}
\usepackage{epstopdf}
\usepackage{pdflscape}
\usepackage{booktabs}
\usepackage{txfonts}
\usepackage{gensymb}
\usepackage{color}
\usepackage{url}
\usepackage{multirow}
\usepackage{amsmath}
\usepackage{amstext}
\usepackage{natbib}
\usepackage{longtable}
\usepackage{float}
\usepackage[export]{adjustbox}

\begin{document}

\title{Near-Infrared Polarimetric Study of Near-Earth Object 252P/LINEAR: An Implication of Scattered Light from the Evolved Dust Particles}

\author{Yuna G. Kwon\inst{\ref{inst1}}~\and~Masateru Ishiguro\inst{\ref{inst1}} \and Jungmi Kwon\inst{\ref{inst2}} \and Daisuke Kuroda\inst{\ref{inst3}} \and Myungshin Im\inst{\ref{inst1}} \and Changsu Choi\inst{\ref{inst1}} \and Motohide Tamura\inst{\ref{inst2},\ref{inst4},\ref{inst5}}  \and Takahiro Nagayama\inst{\ref{inst6}} \and Nobuyuki Kawai\inst{\ref{inst7}} \and Jun-Ichi Watanabe\inst{\ref{inst4}} }

\institute{Department of Physics and Astronomy, Seoul National University, 1 Gwanak, Seoul 08826, Republic of Korea\\
\email{ynkwon@astro.snu.ac.kr}\label{inst1} 
\and Department of Astronomy, Graduate School of Science, The University of Tokyo, 7-3-1 Hongo, Bunkyo-ku, Tokyo 113-0033, Japan\label{inst2}
\and Okayama Observatory, Kyoto University, 3037-5 Honjo, Kamogata, Asakuchi, Okayama 719-0232, Japan\label{inst3}
\and National Astronomical Observatory of Japan, 2-21-1 Osawa, Mitaka, Tokyo 181-8588, Japan\label{inst4}
\and Astrobiology Center, 2-21-1 Osawa, Mitaka, Tokyo 181-8588, Japan\label{inst5}
\and Faculty of Science, Kagoshima University, 21-24 Korimoto, Kagoshima, Kagoshima 890-8580, Japan\label{inst6}
\and Department of Physics, Tokyo Institute of Technology, Meguro, Tokyo 152-8551, Japan\label{inst7}}

\date{Received March 26, 2019 / Accepted July 8, 2019}

\abstract {Comets have been resurfacing under solar radiation while preserving their primordial interiors. Multi-epoch observations of comets enable us to characterize a change in sublimation pattern as a function of heliocentric distance, which in turn provides information on the dust environment of comets. }
{We aim to constrain the size and porosity of ejected dust particles from comet 252P/LINEAR and their evolution near the perihelion via  near-infrared (NIR) multiband polarimetry. A close approach of the comet to the Earth in March 2016 ($\sim$0.036 au) provided a rare opportunity for the sampling of the comet with a high spatial resolution.}
{We made NIR $J$$H$$K_{\rm S}$ bands (1.25--2.25 $\mu$m) polarimetric observations of the comet for 12 days near perihelion, interspersed between broadband optical (0.48--0.80 $\mu$m) imaging observations over four months. In addition, dynamical simulation of the comet was performed 1000 yr backward in time.} 
{We detected two discontinuous brightness enhancements of 252P/LINEAR. Before the first enhancement, the NIR polarization degrees of the comet were far lower than those of ordinary comets at a given phase angle. Soon after the activation, however, they increased by $\sim$13 \% at most, showing unusual blue polarimetric color over the $J$ and $H$ bands ($-$2.55 \% $\mu$m$^{\rm -1}$ on average) and bluing of both $J-H$ and $H-K_{\rm S}$ dust color. Throughout the event, the polarization vector was marginally aligned perpendicular to the scattering plane (i.e., $\theta_{\rm r}$ = 4.6\degree--10.9\degree). The subsequent postperihelion reactivation of the comet lasted for approximately 1.5 months, with a factor of $\sim$30 times pre-activation dust mass-loss rates in the $R_{\rm C}$ band.} 
{ The marked increase in the polarization degree with blue NIR polarimetric color is reminiscent of the behaviors of a fragmenting comet D/1999 S4 (LINEAR). The most plausible scenario for the observed polarimetric properties of 252P/LINEAR would be an ejection of predominantly large (well in geometrical optics regime), compact dust particles from the desiccated surface layer. We conjecture that the more intense solar heating that the comet has received in the near-Earth orbit would cause the paucity of small, fluffy dust particles around the nucleus of the comet.}

\keywords{Comets: general -- Comets: individual: 252P/LINEAR -- Methods: observational -- Techniques: polarimetric, photometric}

\titlerunning{Activations of 252P/LINEAR during its 2016 apparition}

\authorrunning{Y. G. Kwon et al.}

\maketitle

\section{Introduction \label{sec:intro}}

Comets, which are among the least-altered leftovers from the early solar system, have probably preserved their primitive structures inside, whereas their surfaces become different from initial states after repetitive orbital revolutions around the Sun. Solar radiation depletes near-surface volatiles and grains with high surface-to-volume ratios \citep{Prialnik2004}, while concurrent sintering effects would lead to the growth of grain-to-grain contact areas and condensation of them \citep{Kossacki1994,Thomas1994,Gundlach2018}. As a result, resurfacing by the inert refractory layer (the so-called `dust mantle') makes the surface drier and more consolidated than the bulk nuclei \citep{Biele2015,Spohn2015,Kossacki2015}. The cometary dust that we observe generally originates from such a near-surface layer or bounded boulders ejected from the last apparition \citep{Rotundi2015,Fulle2018}. Conversely, fresh particles tend to be ejected from the interiors of the nuclei in rather erratic ways, such as during a sudden brightness enhancement near perihelion or outbursts (e.g., \citealt{Ishiguro2010,Ishiguro2016}).

The polarimetry of comets is a useful diagnostic for investigating the physical properties of dust particles, such as their sizes and porosities \citep{Kiselev2015}. In particular, polarimetric observations at near-infrared wavelengths (NIR; 1.2 $\mu$m--2.3 $\mu$m) have been predicted to set constraints on the porosity of dust grains by covering more dust monomers (basic constitutional units of dust aggregates with $\sim$0.1 $\mu$m radii; e.g., \citealt{Kimura2006}) within a single wavelength than in the optical \citep{Kolokolova2010}. Electromagnetic  interaction between the monomers in aggregates depolarizes the light, randomizing directional information of the scattered light, so that the lower the porosity of a dust particle is, the greater the depolarization as the wavelength increases. It is thus theoretically expected that NIR polarimetry would maximize the difference between the porosities of cometary dust, showing a more distinct contrast between the evolved, hardened particles and fresh, likely fluffier dust particles compared to the optical polarimetry \citep{Kolokolova2010}. As of now, a dozen comets have been polarimetrically observed at NIR, with a bias toward dynamically new or long-periodic comets. Three of them were observed with multiple NIR filters (quasi-)simultaneously (i.e., comets C/1995 O1 (Hale-Bopp), 1P/Halley, and C/1975 V1 (West)), with the last one being the only comet observed at large phase angles of $\alpha$ > 70\degr\ (\citealt{Kiselev2015} and references therein). 

Here, we present the results of tracking the evolution of the activity of comet 252P/LINEAR (hereafter ``252P'') over four months in 2016, with multiband NIR imaging polarimetric and optical imaging observations. 252P attracted attention due to its close approach to the Earth ($\sim$0.036 au) and its possible dynamical pairing with comet P/2016 BA14 (PanSTARRS). Additionally, 252P had been one of the weakly active comets \citep{Ye2016a}, despite its recent ejection into the near-Earth orbit ($q$ < 1.3 au, where $q$ denotes the perihelion distance), which led \citet{Ye2016a, Ye2016b} to suggest that 252P probably has a volatile-poor origin. However, we questioned such a scenario for the comet based on (i) the strong C$_{\rm 2}$-rich coma of 252P observed in 2016 \citep{McKay2017}, (ii) observational evidence suggesting the existence of icy chunks near the nucleus of the comet \citep{Li2017,Coulson2017}, and (iii) an estimated elapsed time of 252P in the near-Earth orbit ($\sim$3 $\times$ 10$^{\rm 2}$ yr; \citealt{Tancredi2014}), which could suffice for the growth of the dust mantle ($\sim$1--10 yr near the Sun; \citealt{Jewitt2002,Kwon2016}).

What if the long-lasting dormancy of 252P might originate in insulation by the sturdy mantle, not in a lack of volatiles? How does the light scattered by such evolved dust particles behave in comparison with the light scattered by freshly ejected particles from the interior? Our science objective is to characterize the activity of this near-Earth comet in its 2016 apparition. The favorable observational geometry of the comet in 2016 enabled us to sample the bright short-periodic comet with a high spatial resolution using ground-based telescopes. 

This paper is divided as follows. Section \ref{sec:obsdata} describes the observational methods and data analyses, and Section \ref{sec:res1} presents the results. Compared with the optical imaging observations conducted over $\sim$four months, the NIR polarimetric observations only overlap for a short period of $\sim$two weeks before the perihelion passage. Hence in Section \ref{sec:res1}, we cope with the observational results of the data both taken at pre-perihelion, which are discussed in Section \ref{sec:discuss}. Finally, Section \ref{sec:sum} presents a summary. The collateral post-perihelion photometric results are present in Appendix \ref{sec:postphot}, showing the optical colorimetric results favorable for the presence of icy particles, and the results of backward dynamical simulation and discussion on the size of the comet are described in Appendices \ref{sec:orbit} and \ref{sec:size}, respectively.

\begin{table*}[!t]
\centering
\caption{Observational geometry and instrument settings}
\vskip-1ex
\begin{tabular}{c|c|c|c|cccccc}
\toprule
\multirow{2}{*}{\it Mode} & \multirow{2}{*}{Telescope/Instrument} & Median UT & \multirow{2}{*}{Filter} & \multirow{2}{*}{\it N} & \multirow{2}{*}{\it Exptime} & \multirow{2}{*}{$r_{\rm H}$} & \multirow{2}{*}{$\Delta$} & \multirow{2}{*}{$\alpha$} & \multirow{2}{*}{$\nu$}\\
& & 2016+ & & &  & & & & \\
\midrule
\midrule
\multirow{9}{*}{\it Image} & \multirow{9}{*}{LSGT/SBIG ST-10} & 02/13 11:00 & \multirow{9}{*}{$R$} & 90 & 30.0 & 1.085 & 0.215 & 57.7 & 322.7 \\
  & & 02/22 10:53 & & 100 & 30.0 & 1.042 & 0.168 & 67.1 & 332.8 \\
  & & 02/24 10:48 & & 50 & 60.0 & 1.034 & 1.057 & 69.2 & 335.2 \\
  & & 02/26 10:38 & & 50 & 60.0 & 1.027 & 0.146 & 71.3 & 337.6 \\
  & & 02/28 11:12 & & 25 & 60.0 & 1.021 & 0.135 & 73.4 & 340.0 \\
  & & 03/01 10:44 & & 50 & 60.0 & 1.015 & 0.125 & 75.4 & 342.5 \\
  & & 03/05 10:39 & & 100 & 30.0 & 1.006 & 0.103 & 79.4 & 347.5 \\
  & & 03/07 10:44 & & 65 & 30.0 & 1.002 & 0.092 & 81.3 & 350.0 \\
  & & 03/09 11:04 & & 100 & 5.0 & 0.999 & 0.081 & 83.1 & 352.6 \\
\midrule
  \multirow{8}{*}{\it Impol} & \multirow{8}{*}{IRSF/SIRPOL} & 03/04 20:16 & \multirow{8}{*}{$JHK_{\rm S}$} & 60 & 20.0 & 1.007 & 0.106 & 78.8 & 346.7 \\
  & & 03/09 20:27 & & 60 & 10.0 & 0.999 & 0.079 & 83.5 & 353.1 \\
  & & 03/11 18:12 & & 20 & 5.0 & 0.997 & 0.069 & 85.0 & 355.5 \\
  & & 03/12 21:19 & & 100 & 5.0 & 0.997 & 0.063 & 85.8 & 356.9 \\
  & & 03/13 19:43 & & 100 & 5.0 & 0.996 & 0.059 & 86.3 & 358.1 \\
  & & 03/14 18:49 & & 100 & 5.0 & 0.996 & 0.054 & 86.7 & 359.4 \\
  & (Perihelion) & (03/15 06:38) & & (\ldots) & (\ldots) & (0.996) & (0.052) & (86.8) & (0.0) \\
  & & 03/15 19:56 & & 100 & 5.0 & 0.996 & 0.050 & 87.0 & 0.7 \\
\midrule
  \multirow{27}{*}{\it Image} & \multirow{27}{*}{OAO-50/MITSuME} & 03/24 19:47 & \multirow{27}{*}{g$'R_{\rm C}I_{\rm C}$} & 12 & 30.0 & 1.005 & 0.041 & 77.6 & 12.2 \\
  & & 03/25 19:31 & & 40 & 60.0 & 1.007 & 0.044 & 76.1 & 13.4 \\
  & & 03/26 19:20 & & 50 & 60.0 & 1.009 & 0.048 & 74.8 & 14.7 \\
  & & 03/27 18:53 & & 100 & 60.0 & 1.012 & 0.052 & 73.5 & 15.9 \\
  & & 03/28 18:55 & & 95 & 60.0 & 1.014 & 0.056 & 72.2 & 17.1 \\
  & & 04/05 19:30 & & 35 & 60.0 & 1.041 & 0.098 & 63.3 & 26.8 \\
  & & 04/11 18:09 & & 145 & 60.0 & 1.068 & 0.132 & 57.1 & 33.6 \\
  & & 04/15 17:24 & & 150 & 60.0 & 1.089 & 0.154 & 53.0 & 38.0 \\
  & & 04/17 17:15 & & 65 & 60.0 & 1.100 & 0.166 & 51.0 & 40.1 \\
  & & 04/18 17:44 & & 180 & 60.0 & 1.106 & 0.171 & 50.0 & 41.2 \\
  & & 04/19 16:49 & & 80 & 60.0 & 1.112 & 0.177 & 49.0 & 42.2 \\
  & & 04/22 17:02 & & 30 & 120.0 & 1.130 & 0.194 & 46.1 & 45.2 \\
  & & 04/24 17:38 & & 65 & 120.0 & 1.143 & 0.206 & 44.2 & 47.2 \\
  & & 04/25 16:59 & & 80 & 120.0 & 1.150 & 0.212 & 43.3 & 48.1 \\
  & & 04/28 16:31 & & 50 & 120.0 & 1.170 & 0.229 & 40.5 & 51.0 \\
  & & 04/29 17:18 & & 75 & 120.0 & 1.178 & 0.235 & 39.6 & 51.9 \\
  & & 04/30 17:14 & & 85 & 120.0 & 1.185 & 0.241 & 38.7 & 52.8 \\
  & & 05/02 16:56 & & 30 & 120.0 & 1.199 & 0.253 & 37.0 & 54.6 \\
  & & 05/04 17:15 & & 60 & 120.0 & 1.214 & 0.266 & 35.2 & 56.4 \\
  & & 05/11 17:15 & & 70 & 120.0 & 1.268 & 0.310 & 29.8 & 62.1 \\
  & & 05/12 17:13 & & 70 & 120.0 & 1.276 & 0.317 & 29.1 & 62.9 \\
  & & 05/14 17:17 & & 45 & 120.0 & 1.292 & 0.331 & 27.7 & 64.5 \\
  & & 05/17 17:19 & & 20 & 120.0 & 1.316 & 0.352 & 25.8 & 66.7 \\
  & & 05/18 16:34 & & 80 & 120.0 & 1.324 & 0.359 & 25.2 & 67.4 \\
  & & 05/20 16:37 & & 75 & 120.0 & 1.341 & 0.373 & 24.2 & 68.9 \\
  & & 06/02 18:00 & & 9 & 120.0 & 1.453 & 0.480 & 19.6 & 77.4 \\
  & & 06/10 17:58 & & 15 & 120.0 & 1.524 & 0.555 & 19.1 & 82.0\\
\bottomrule
\end{tabular}
\tablefoot{Top headers: $Mode$, instrumental settings of imaging ($Image$) and imaging-polarimetry ($Impol$) observations; $N$, number of exposures; $Exptime$, individual exposure time in seconds; $r_{\rm H}$ and $\Delta$, median heliocentric and geocentric distances in au, respectively; $\alpha$, median phase angle (angle of Sun--comet--observer) in degrees; and $\nu$, median true anomaly in degrees. We referred to the web-based JPL Horizons system (http://ssd.jpl.nasa.gov/?horizons) to obtain the ephemerides.}
\label{t1}
\vskip-1ex
\end{table*}

\begin{table*}[!t]
\centering
\caption{Nightly averaged photometric and polarimetric parameters of 252P/LINEAR taken by IRSF/SIRPOL}
\vskip-1ex
\begin{tabular}{c|c|cc|cc|ccc|cc}
\toprule
UT 2016+ & Filter & $P$ $^{\rm a}$ & $\sigma_{\rm P}$ $^{\rm b}$ & $\theta_{\rm P}$ $^{\rm c}$ & $\sigma_{\theta_{\rm P}}$ $^{\rm d}$ & $P_{\rm r}$ $^{\rm e}$ & $\sigma_{P_{\rm r}}$ $^{\rm f}$ & $\theta_{\rm r}$ $^{\rm g}$ & $m_{\rm IRSF}$ $^{\rm h}$ \\
\midrule
\midrule
\addlinespace[0.15cm]
 MAR 04 & $J$ & 19.46 & 0.85 & $-$0.68 & 1.01 & 18.12 & 1.41 & 10.69 & 13.28 $\pm$ 0.19 \\
   ($\phi$$^{\rm i}$ = 78.7\degree) & $H$ & 19.57 & 0.79 & $-$0.46 & 0.47 & 18.17 & 0.92 & 10.91 & 12.60 $\pm$ 0.18 \\
\midrule
  \multirow{2}{*}{MAR 09} & $J$ & 26.04 & 0.99 & $-$0.34 & 0.29 & 25.19 & 0.99 & 7.33 & 12.49 $\pm$ 0.17 \\
  & $H$ & 24.00 & 0.91 & $-$0.38 & 0.18 & 23.23 & 0.90 & 7.28 & 11.80 $\pm$ 0.11 \\
  ($\phi$ = 82.4\degree) & $K_{\rm S}$ & 24.05 & 0.96 & $-$0.24 & 0.54 & 23.25 & 1.04 & 7.43 & 11.60 $\pm$ 0.15 \\
\midrule
  \multirow{2}{*}{MAR 11} & $J$ & 28.19 & 1.30 & $-$0.43 & 0.37 & 27.61 & 1.31 & 5.79 & 12.74 $\pm$ 0.12 \\
  & $H$ & 27.88 & 1.13 & $-$0.37 & 0.10 & 27.31 & 1.11 & 5.85 & 12.24 $\pm$ 0.10 \\
  ($\phi$ = 83.9\degree) & $K_{\rm S}$ & 27.74 & 1.56 & $-$0.15 & 1.85 & 27.12 & 2.11 & 6.10 & 12.08 $\pm$ 0.20 \\
\midrule
  \multirow{2}{*}{MAR 12} & $J$ & 32.57 & 1.22 & 0.27 & 0.39 & 31.92 & 1.23 & 5.73 & 11.13 $\pm$ 0.11 \\
  & $H$ & 31.69 & 1.14 & 0.23 & 0.36 & 31.07 & 1.16 & 5.69 & 11.20 $\pm$ 0.11 \\
  ($\phi$ = 84.5\degree) & $K_{\rm S}$ & 27.35 & 1.06 & 0.24 & 0.51 & 26.81 & 1.10 & 5.70 & 11.58 $\pm$ 0.12 \\
\midrule
  \multirow{2}{*}{MAR 13} & $J$ & 29.60 & 1.11 & 0.21 & 0.44 & 29.13 & 1.13 & 5.13 & 11.12 $\pm$ 0.13 \\
  & $H$ & 28.73 & 1.03 & 0.21 & 0.36 & 28.27 & 1.04 & 5.13 & 11.22 $\pm$ 0.10 \\
  ($\phi$ = 85.2\degree) & $K_{\rm S}$ & 28.36 & 1.06 & 0.18 & 0.46 & 27.92 & 1.09 & 5.10 & 11.29 $\pm$ 0.11 \\
\midrule
  \multirow{2}{*}{MAR 14} & $J$ & 27.37 & 1.02 & 0.21 & 0.46 & 27.02 & 1.05 & 4.64 & 11.86 $\pm$ 0.17 \\
  & $H$ & 26.03 & 0.94 & 0.23 & 0.36 & 25.69 & 0.95 & 4.66 & 11.41 $\pm$ 0.10 \\
  ($\phi$ = 85.6\degree) & $K_{\rm S}$ & 26.19 & 0.98 & 0.19 & 0.56 & 25.85 & 1.02 & 4.62 & 11.27 $\pm$ 0.20 \\
\midrule
  \multirow{2}{*}{MAR 15} & $J$ & 28.10 & 1.11 & 0.19 & 0.59 & 27.73 & 1.15 & 4.68 & 11.84 $\pm$ 0.12 \\
  & $H$ & 27.28 & 1.01 & 0.26 & 0.38 & 26.91 & 1.02 & 4.75 & 11.37 $\pm$ 0.10 \\
  ($\phi$ = 85.9\degr) & $K_{\rm S}$ & 26.87 & 1.05 & 0.29 & 0.46 & 26.49 & 1.08 & 4.77 & 11.25 $\pm$ 0.19 \\
\bottomrule
\end{tabular}
\tablefoot{
\tablefoottext{\rm a}{ Observed degree of linear polarization of the comet in percent;}
\tablefoottext{\rm b}{ Standard deviation of $P$ in percent;}
\tablefoottext{\rm c}{ Electric vector position angle in degrees;}
\tablefoottext{\rm d}{ Standard deviation of $\theta_{\rm P}$ in degrees;}
\tablefoottext{\rm e}{ $P$ relative to the scattering plane in percent (see Eq. \ref{eq:eq5});}
\tablefoottext{\rm f}{ Standard deviation of $P_{\rm r}$ in percent;}
\tablefoottext{\rm g}{ Polarization angle with regard to the perpendicular direction of the scattering plane in degrees (see Eq. \ref{eq:eq6});}
\tablefoottext{\rm h}{ Apparent magnitude of the IRSF data;}
and \tablefoottext{\rm i}{ Position angle of the scattering plane.}}
\label{t2}
\vskip-1ex
\end{table*}

\section{Observations and data analyses \label{sec:obsdata}}

\subsection{Observations \label{sec:obs}}

The journal of our observational geometry and instrument settings is shown in Table \ref{t1}. NIR imaging polarimetric observations were conducted from UT 2016 March 02 to March 15 in daily cadence using a polarimeter SIRPOL of the NIR camera SIRIUS attached to the InfraRed Survey Facility (IRSF) 1.4 m diameter telescope at the South African Astronomical Observatory (32\degr22$\arcmin$46$\arcsec$E, $-$20\degr48$\arcmin$39$\arcsec$S, 1798 m), Sutherland, South Africa. SIRPOL, a single-beam polarimeter of a half-wave plate (HWP) rotator with a wire grid polarizer located upstream of the SIRIUS camera at the Cassegrain focus, simultaneously provides multiband NIR wide-field polarimetry (7.7\arcmin $\times$ 7.7\arcmin, with the image scale of 0.45\arcsec pixel$^{\rm -1}$) at the $J$ (center wavelength $\lambda_{\rm C}$ = 1.25 $\mu$m), $H$ ($\lambda_{\rm C}$ = 1.65 $\mu$m), and $K_{\rm S}$ ($\lambda_{\rm C}$ = 2.25 $\mu$m) bands \citep{Nagayama2003,Kandori2006}. A microcontroller controls the HWP rotator angle by driving a stepping motor and takes images in the sequence of 0\degr, 45\degr, 22.5\degree, and 67.5\degr. Because of the unavailability of the nonsidereal tracking mode, we set the exposure time of 5--20 sec so that the elongation of the comet was smaller than $\sim$2\arcsec. Target observations were conducted in units of 10 sets, each of which consists of exposures at four HWP angles (0\degr, 45\degr, 22.5\degree, and 67.5\degree) dithered by $\sim$30 pixels (13.5\arcsec, which was large enough to avoid the coma signals). We interleaved sky observations $+$400 pixels in x and $-$400 pixels in y directions ($\pm$3$\arcmin$) off the detector center with every 10 sets of target observations. Seeing ranged from  0.9\arcsec--1.8\arcsec and was primarily 1.2\arcsec.  

Preperihelion Johnson $R$ band ($\lambda_{\rm C}$ = 0.62 $\mu$m) imaging observations were obtained from UT 2016 February 13 to March 09 using the 0.4 m diameter Lee Sang Gak telescope (LSGT) at the Siding Spring Observatory (149\degr03$\arcmin$52$\arcsec$E, $-$31\degr16$\arcmin$24$\arcsec$S, 1165m), New South Wales, Australia. We used the SBIG ST-10 camera, the CCD chip of which has 2184 $\times$ 1472 pixels with a 6.8 $\mu$m pixel pitch. It provides an image scale of 0.48\arcsec with a field of view of 17.5\arcmin $\times$ 11.8\arcmin \citep{Im2015}. Because of the unavailability of the nonsidereal tracking mode, we set the exposure time of 5--60 sec so that the elongation of the comet was smaller than $\sim$1\arcsec. Seeing was approximately  1.2\arcsec. 

Postperihelion simultaneous multiband (Sloan Digital Sky Survey (SDSS) g$'$ and Johnson-Cousins $R_{\rm C}$ and $I_{\rm C}$ bands, the $\lambda_{\rm C}$ of which are 0.48, 0.66, and 0.80 $\mu$m, respectively) imaging observations were performed from UT 2016 March 24 to June 10 using the 0.5 m diameter telescope at the Okayama Astrophysical Observatory (OAO-50; 133\degr35$\arcmin$36$\arcsec$E, $+$34\degr34$\arcmin$33$\arcsec$N, 360 m), Okayama, Japan. We employed the Multicolor Imaging Telescopes for Survey and Monstrous Explosions (MITSuME) system, which consists of three 1024 $\times$ 1024 CCD chips with a 24.0 $\mu$m pixel pitch \citep{Kotani2005}. It covers the field of view of 26$\arcmin$ $\times$ 26$\arcmin$ with a pixel resolution of  1.53\arcsec. Nonsidereal guiding was conducted so that we obtained integrations of 30--120 sec, each depending on the signal-to-noise ratio (SNR) of the comet. Airmass was variable in the range 1.1\arcsec--5.1\arcsec and seeing was primarily 2.5\arcsec.

\subsection{Data analyses  \label{sec:data}}

The raw observational (polarimetric and photometric) data were preprocessed with standard techniques for imaging data: bias and dark subtractions, flat-fielding, frame registration, and sky subtraction in IRAF. The pixel coordinates of the images were converted into celestial coordinates using WCSTools \citep{Mink1997}, with field stars listed in the 2MASS catalog \citep{Skrutskie2006} for the IRSF data and with stars listed in the UCAC-3 catalog \citep{Zacharias2010} for the LSGT and OAO data. Since incomplete sky subtraction can produce a spurious false degree of linear polarization ($P$) signals at NIR, we took special care for the background subtraction at the level of each HWP angle image. Each set of sky frames was median-combined by matching the stars with identical wcs information and was subtracted from the target images taken at the same HWP angle. If there are still remaining counts after reduction by the standard sky subtraction process, we subtracted an arbitrary constant value measured at areas of blank sky well outside the coma to make the sky count nearly zero. The resulting object images were median-combined nightly by matching the instantaneous location of the comet as the center to eliminate contaminations (e.g., noise from background stars and cosmic rays) and to improve the SNR. In this process, we used data only if the SNR of the comet in a single image exceeds three. 

The $P$ of the comet was derived by performing photometry of focused images using APPHOT in IRAF. Since the SNR of the resulting combined images ($\lesssim$ 80) is not high enough to perform imaging polarimetry, we decided to perform aperture polarimetry. Therefore, we integrated all signals within an aperture, losing the spatial information inside. We implemented a photopolarimetric aperture of 5 pixels (2.25\arcsec, corresponding to the projected physical distance of 82--173 km during the polarimetric observation) in radius in all ($J$, $H$, and $K_{\rm S}$) bands. The  Stokes $I$ and, subsequently, Stokes $Q$ and $U$ were calculated using 
\begin{equation}
 I = \frac{(I_{\rm 0} + I_{\rm 45} + I_{\rm 22.5} + I_{\rm 67.5})}{2}~, 
\label{eq:eq1}
\end{equation} 
and
\begin{equation}
 Q = I_{\rm 0} - I_{\rm 45}, ~~~~~~~U = I_{\rm 22.5} - I_{\rm 67.5} ~,
\label{eq:eq2}
\end{equation}
where $I_{\rm X}$ denotes the intensity taken at a HWP angle of X in degrees. The $P$ and the polarization position angle of the strongest electric vector $\theta_{\rm P}$ can be calculated using 
\begin{equation}
 P = \frac{\sqrt{Q^{\rm 2} + U^{\rm 2}}}{I}~, 
\label{eq:eq3}
\end{equation} 
and
\begin{equation}
 \theta_{\rm P} = \frac{1}{2}~{\rm arctan}~\biggl(\frac{U}{Q}\biggl)~.
\label{eq:eq4}
\end{equation}
The above quantities ought to be corrected for instrumental effects, that is, for internal polarization, polarization efficiencies of each filter and position angle offset of SIRPOL. The polarization efficiencies of the $J$, $H$, and $K_{\rm S}$ filters are known to be 95.5 \%, 96.3 \%, and 98.5 \%, respectively, and the polarization position angle offset of the instrument is 105\degr\ in the bands \citep{Kandori2006}. For instrumental polarization ($P_{\rm inst}$), as mentioned in detail in previous studies using SIRPOL (e.g., \citealt{Kandori2006,Kwon2015}) as well as in the recently updated results \citep{Kusune2015}, the $P_{\rm inst}$ of SIRPOL has been considered to be negligible. Since 252P was observed at high phase angles ($\alpha$ = 78.8\degree--87.0\degree), where the majority of comets exhibit $P$ $\gtrsim$ 20 \% \citep{Kiselev2015}, an influence of $P_{\rm inst}$ on our results should be inconsequential.

Finally, the corrected $P$ and $\theta_{\rm P}$ were converted into quantities with respect to the scattering plane (the plane on which the Sun--comet--Earth are located) as follows:
\begin{equation}
 P_{\rm r} = P~{\rm cos}~(2 \theta_{\rm r})~,
\label{eq:eq5}
\end{equation}
and
\begin{equation}
 \theta_{\rm r} = \theta_{\rm P} - \biggl(\phi \pm \frac{\pi}{2}\biggl)~,
\label{eq:eq6}
\end{equation}
where $\phi$ represents the position angle of the scattering plane, the sign of which ($\pm$ in Eq. \ref{eq:eq6}) was chosen to satisfy 0 $\le$ ($\phi$ $\pm$ $\pi$/2) $\le$ $\pi$ \citep{Chernova1993}. Uncertainties were estimated in a standard error propagation law. We tabulated the resulting polarimetric parameters and their errors in Table \ref{t2}. For polarimetric analyses, we excluded all data taken on UT 2016 March 02--03 and the data in the $K_{\rm S}$ band on March 04 because of the faint comet brightness (SNR $\lesssim$ 3) and temporal malfunction of the $K_{\rm S}$ band detector, respectively, and the data on March 05--08 because of high and fluctuating humidity conditions.

Photometric aperture sizes of the imaging (LSGT and OAO) and Stokes $I$ (IRSF) data were all restricted to the projected physical distance of 1000 km from the center of the nucleus (corresponding to 13--35 pixels for the LSGT data, 2--22 pixels for the OAO data, and 29--61 pixels for  the IRSF data during the observing epochs) in radius. Sky brightness was subtracted by the circular annuli with widths of 5 pixels just outside the employed aperture. Flux calibration of the g$'$$R_{\rm C}I_{\rm C}$ filters was obtained from images of nearby comparison stars listed in the AAVSO Photometric All Sky Survey DR9 catalog (APASS; \citealt{Henden2016}). We assumed a systematic error of the catalog of $\sim$0.1 mag to take into account the calibration uncertainties \citep{Tonry2018}. SDSS magnitudes of the APASS catalog were transformed to the Johnson-Cousins magnitudes using the prescriptions for the main-sequence stars by \citet{Jester2005} and \citet{Jordi2005}. In addition, we converted the IRSF magnitude system, which follows the Mauna Kea Observatories Near-Infrared filter system \citep{Tokunaga2002}, into the Johnson-Cousins $R_{\rm C}$ magnitudes to know the approximate trend of the cometary activity during the IRSF observations compared to the optical data. We analyzed the Stokes $I$ of the $H$-band data as a representative, simply due to the highest SNR of the comet therein. Flux calibration of the IRSF magnitudes was obtained from the field stars of the 2MASS Point Source Catalog \citep{Curti2003} using the conversion equations of \citet{Kato2007}. The $H$ magnitudes were then converted into the Johnson $R$ magnitudes, assuming a solar-like spectrum of the comet for simplicity (i.e., $R-H$ = 1.055; \citealt{Binney1998}), and finally converted into the $R_{\rm C}$ magnitudes using the prescriptions of $R$ $-$ $R_{\rm C}$ = $-$0.17 \citep{Fernie1983}. We tabulated the apparent magnitudes of the IRSF data in Table \ref{t2}.

\begin{figure*}[!t]
\centering
\includegraphics[width=13cm]{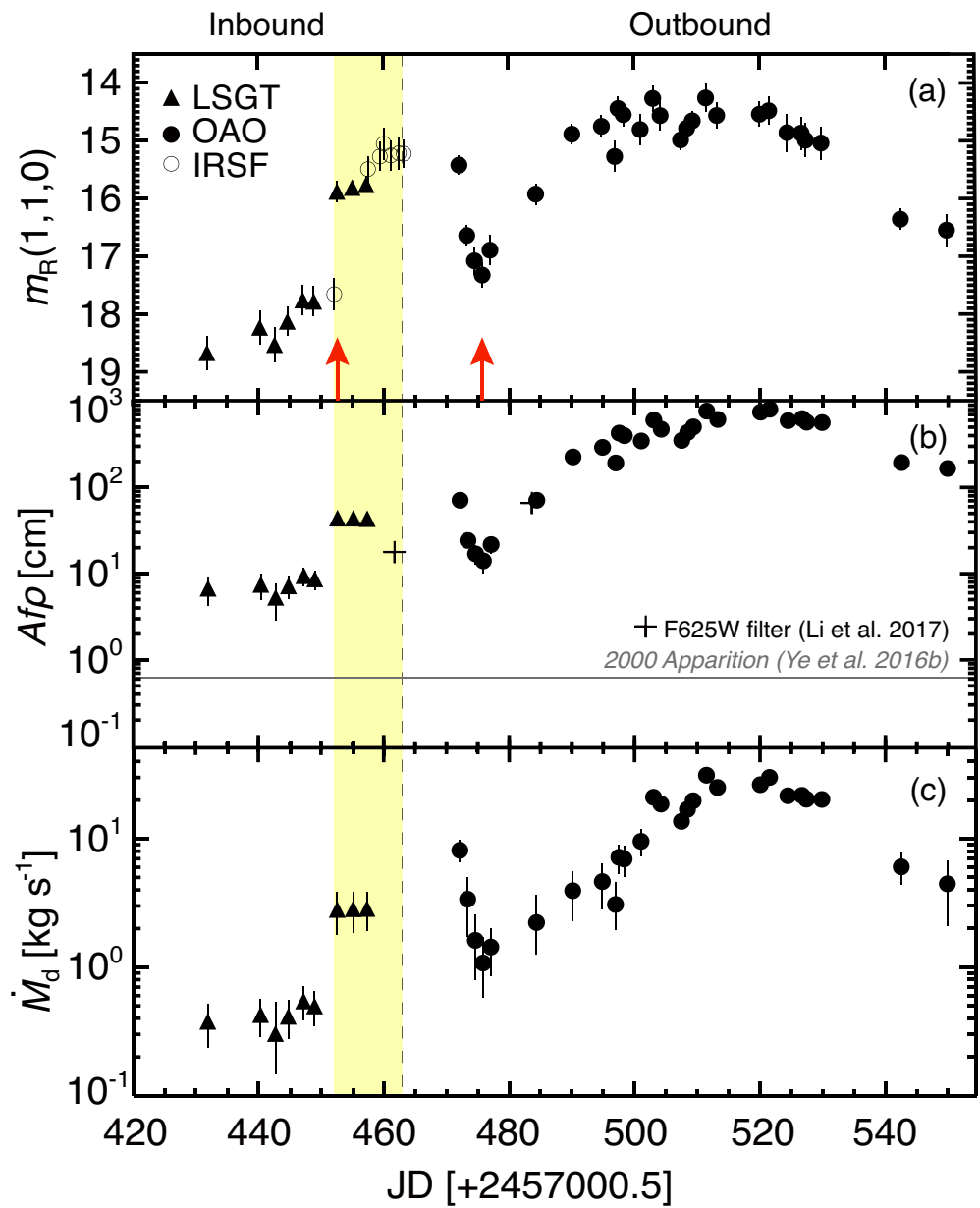}
\caption{(a) Absolute magnitudes $m_{\rm R}$(1, 1, 0), (b) $Af\rho$, and (c) dust mass-loss rate $\dot{M}_{\rm d}$ of 252P observed from UT 2016 February 13 to June 10 by LSGT/SBIG ST-10 (filled triangles), OAO-50/MITSuME (filled circles), and IRSF/SIRPOL (open circles) are shown as a function of the Julian Date (JD). The dashed vertical lines denote the perihelion at $r_{\rm H}$ = 0.996 au on UT 2016 March 15.28. Two arrows in panel (a) mark presumable locations where 252P presented significant brightness enhancements. In panel (b), the horizontal line of $Af\rho$ = 0.6 denotes the median $Af\rho$ value of the comet in the 2000 apparition \citep{Ye2016b}. Two crosses are quoted from the Hubble Space Telescope data of \citet{Li2017} observed within the aperture radius of $\lesssim$10 km from the center in the F625W filter. Yellow-colored area covers the epoch of our polarimetric observations.}
\label{Fig1}
\vskip-1ex
\end{figure*}

\section{Results \label{sec:res1}}

\begin{table*}[!t]
\centering
\caption{Photometry of 252P/LINEAR}
\begin{tabular}{c|c|ccccccc}
\toprule
\multirow{2}{*}{Telescope} & Median UT & \multirow{2}{*}{$\varrho$ [$\arcsec$]} & & \multirow{2}{*}{$m_{\rm R}$(1, 1, 0) [mag]} & & \multirow{2}{*}{$Af\rho$ [cm]} && \multirow{2}{*}{$\dot{M}_{\rm d}$ [kg s$^{\rm -1}$]} \\
& 2016+ & & & & & & &  \\
\midrule
\midrule
\multirow{9}{*}{LSGT} & 02/13 11:00 & 6.41 && 18.67 $\pm$ 0.30 && 6.78 $\pm$ 2.50 && 0.38 $\pm$ 0.14\\
& 02/22 10:53 & 8.21 && 18.23 $\pm$ 0.30 && 7.51 $\pm$ 2.50 && 0.43 $\pm$ 0.14\\
& 02/24 10:48 & 8.78 && 18.53 $\pm$ 0.30 && 5.33 $\pm$ 2.45 && 0.30 $\pm$ 0.23\\
& 02/26 10:38 & 9.43 && 18.13 $\pm$ 0.24 && 7.23 $\pm$ 2.20 && 0.41 $\pm$ 0.14\\
& 02/28 11:12 & 10.19 && 17.76 $\pm$ 0.25 && 9.48 $\pm$ 2.25 && 0.55 $\pm$ 0.16\\
& 03/01 10:44 & 11.07 && 17.78 $\pm$ 0.25 && 8.70 $\pm$ 2.20 && 0.50 $\pm$ 0.15\\
& 03/05 10:39 & 13.44 && 15.88 $\pm$ 0.18 && 44.14 $\pm$ 1.48  && 2.85 $\pm$ 1.02\\
& 03/07 10:44 & 15.03 && 15.81 $\pm$ 0.14 && 44.10 $\pm$ 1.25 && 2.82 $\pm$ 1.00\\
& 03/09 11:04 & 17.02 && 15.76 $\pm$ 0.14 && 43.56 $\pm$ 1.24 && 2.62 $\pm$ 0.95\\
\midrule
\multirow{27}{*}{OAO-50}& 03/24 19:47 & 34.30 & & 15.42 $\pm$ 0.12 && 71.05 $\pm$ 3.24 && 8.13 $\pm$ 1.61\\
& 03/25 19:31 & 31.96 && 16.64 $\pm$ 0.13 && 24.27 $\pm$ 3.36 && 3.38 $\pm$ 1.65\\
& 03/26 19:20 & 29.30 && 17.08 $\pm$ 0.22 && 16.94 $\pm$ 4.25 && 1.62 $\pm$ 1.12\\
& 03/27 18:53 & 27.05 && 17.33 $\pm$ 0.19 && 14.07 $\pm$ 3.79 && 1.08 $\pm$ 0.97\\
& 03/28 18:55 & 25.11 && 16.90 $\pm$ 0.25 && 21.77 $\pm$ 4.55 && 1.43 $\pm$ 1.27\\
& 04/05 19:30 & 14.35 && 15.92 $\pm$ 0.14 && 71.04 $\pm$ 3.40 && 2.23 $\pm$ 1.69\\
& 04/11 18:09 & 10.65 && 14.89 $\pm$ 0.13 && 225.84 $\pm$ 3.36 && 3.94 $\pm$ 1.65\\
& 04/15 17:24 & 9.13 && 14.75 $\pm$ 0.16 && 291.78 $\pm$ 5.15 && 4.64 $\pm$ 1.79\\
& 04/17 17:15 & 8.47 && 15.27 $\pm$ 0.26 && 192.36 $\pm$ 4.60 && 3.08 $\pm$ 2.30\\
& 04/18 17:44 & 8.22 && 14.44 $\pm$ 0.17 && 427.23 $\pm$ 8.00 && 7.17 $\pm$ 1.83\\
& 04/19 16:49 & 7.95 && 14.55 $\pm$ 0.17 && 399.17 $\pm$ 7.48 && 6.95 $\pm$ 1.86\\
& 04/22 17:02 & 7.25 && 14.81 $\pm$ 0.26 && 346.12 $\pm$ 8.48 && 9.55 $\pm$ 2.30\\
& 04/24 17:38 & 6.83 && 14.27 $\pm$ 0.21 && 604.79 $\pm$ 13.15 && 21.04 $\pm$ 2.07\\
& 04/25 16:59 & 6.63 && 14.57 $\pm$ 0.23 && 473.73 $\pm$ 10.75 && 18.63 $\pm$ 2.14\\
& 04/28 16:31 & 6.14 && 14.99 $\pm$ 0.14 && 351.14 $\pm$ 5.62 && 13.66 $\pm$ 1.68\\
& 04/29 17:18 & 5.98 && 14.79 $\pm$ 0.12 && 435.16 $\pm$ 6.45 && 16.94 $\pm$ 1.60\\
& 04/30 17:14 & 5.84 && 14.66 $\pm$ 0.12 && 504.63 $\pm$ 7.54 && 19.77 $\pm$ 1.59\\
& 05/02 16:56 & 5.56 && 14.26 $\pm$ 0.22 && 768.66 $\pm$ 17.25 && 31.25 $\pm$ 2.09\\
& 05/04 17:15 & 5.29 && 14.56 $\pm$ 0.20 && 614.42 $\pm$ 12.53 && 25.07 $\pm$ 1.98\\
& 05/11 17:15 & 4.54 && 14.54 $\pm$ 0.19 && 747.90 $\pm$ 14.98 && 26.37 $\pm$ 1.93\\
& 05/12 17:13 & 4.44 && 14.48 $\pm$ 0.22 && 808.69 $\pm$ 17.57 && 30.03 $\pm$ 2.09\\
& 05/14 17:17 & 4.25 && 14.86 $\pm$ 0.33 && 596.15 $\pm$ 17.25 && 21.62 $\pm$ 2.66\\
& 05/17 17:19 & 4.00 && 14.87 $\pm$ 0.29 && 628.16 $\pm$ 16.44 && 21.77 $\pm$ 2.44\\
& 05/18 16:34 & 3.92 && 14.99 $\pm$ 0.30 && 571.81 $\pm$ 15.25 && 20.33 $\pm$ 2.54\\
& 05/20 16:37 & 3.77 && 15.04 $\pm$ 0.29 && 566.12 $\pm$ 14.69 && 20.24 $\pm$ 2.44\\
& 06/02 18:00 & 2.93 && 16.36 $\pm$ 0.14 && 194.27 $\pm$ 3.37 && 6.04 $\pm$ 1.68\\
& 06/10 17:58 & 2.53 && 16.55 $\pm$ 0.27 && 165.99 $\pm$ 4.74 && 4.46 $\pm$ 2.35\\
\bottomrule
\end{tabular}
\tablefoot{All $\varrho$ values correspond to the projected physical distance of 1000 km from the nucleus center.}
\label{t3}
\vskip-1ex
\end{table*}

\subsection{Photometric results I: Detection of two discontinuous activations \label{sec:phot1}}

Figure \ref{Fig1} shows the temporal evolutions of (a) the absolute magnitude $m_{\rm R}$(1, 1, 0), (b) $Af\rho$ parameter, and (c) dust mass-loss rate $\dot{M}_{\rm d}$ of 252P as a function of the Julian Date (JD) during the period from UT 2016 February 13 to June 10. Arrows in panel (a) mark two presumable points at which the comet presented discontinuous brightness enhancements. At a glance, the first activation seemed to occur between March 01 and 05 (i.e., between JD 2457448.5 and 2457452.5), which is roughly consistent with the amateur reports of the webpage by S. Yoshida\footnote{http://www.aerith.net/comet/catalog/0252P/2016.html}. 

To investigate a change in the intrinsic brightness of the comet, it is necessary to eliminate the effect of the varying viewing geometry. Thus, we converted the apparent magnitudes into the reduced absolute magnitudes, which correspond to the magnitude at a hypothetical point in space ($r_{\rm H}$ = $\Delta$ = 1 au and $\alpha$ = 0\degr), using
\begin{equation}
 m_{\rm R}(1,~1,~0)=m_{\rm R}(r_{\rm H},~\Delta,~\alpha)-5\log_{\rm 10}(r_{\rm H}\Delta)-2.5\log_{\rm 10} (\Phi(\alpha)),
\label{eq:eq7}
\end{equation}
\noindent where $m_{\rm R}$($r_{\rm H}$, $\Delta$, $\alpha$) is the apparent magnitude of the comet, and $\Phi(\alpha)$ is the phase function of the coma dust. We adopted a commonly used empirical scattering phase function, i.e., 2.5 $\log_{\rm 10}$$(\Phi(\alpha))$ = $b$$\alpha$, where the phase coefficient of $b$ = 0.035 mag deg$^{\rm -1}$ was assumed (see, e.g., \citealt{Lamy2004}, p. 223). Although the errors associated with the Poisson noise of the OAO data were on the order of 0.001--0.01 mag, we added the standard deviations of the field stars for the differential photometry (always $\lesssim$0.2 mag) and the employed photometric error of the APASS catalog ($\sim$0.1 mag) to derive the resulting uncertainties. We then estimated the $Af\rho$ parameters, a proxy of dust production rate of the comet ($A$ is the albedo of dust particles, and $f$ is their packing density within the aperture radius of $\rho$; \citealt{A'Hearn1984}) to compare the activity level in 2016 with that of the 2000 apparition \citep{Ye2016b} from
\begin{equation}
Af\rho = {\rm Y}~\biggl[\frac{\Delta}{{\rm au}}\biggr]^2~\biggl[\frac{r_{\rm H}}{{\rm au}}\biggr]^2~\biggl[\frac{\rho}{{\rm cm}}\biggr]^{-1} \times 2.512^{(m_\odot -~m_{\rm R})},
\label{eq:eq8}
\end{equation}
\noindent in which $m_\odot$ is the $R_{\rm C}$ band magnitude of the Sun at $r_{\rm H}$ = 1 au ($m_\odot$ = $-$27.11; \citealt{Drilling2000}), Y is the unitary transformation factor of 8.95 $\times$ 10$^{\rm 26}$ for distances \citep{Kwon2016}, and $\rho$ is the considered aperture size (10$^{\rm 3}$ km = 10$^{\rm 8}$ cm in this study). The resulting $Af\rho$ values range from 5.3--9.5 cm, with an average of 7.5 cm before the first activation, already showing an $\sim$13 times higher level a month prior to its perihelion passage compared to the 2000 apparition ($\sim$0.6 cm; \citealt{Ye2016b}). Both $m_{\rm R}$(1, 1, 0) and $Af\rho$ values escalated sharply between March 01 and 05 by $\sim$2 mag and by $\sim$35 cm, respectively, implying a sudden increase in the number density of dust particles within the coma encircled by the aperture radius of 1000 km.

A lower limit of the dust mass-loss rate can be estimated from the optical photometry, using an equation in \citet{Luu1992}:
\begin{equation}
\dot{M}_{\rm d} = \frac{1.1 \times 10^{-3} \pi \rho_{\rm d} \bar{a} \eta r_{\rm obj}^{2}}{\varrho r_{\rm h}^{1/2} \Delta},
\label{eq:eq9}
\end{equation}
\noindent where $\rho_{\rm d}$ is the mass density of dust particles (nominal 1000 kg m$^{\rm -3}$ was assumed), $\bar{a}$ is the average of small particle sizes (1 $\mu$m was assumed; the radius of the most effective scatterers in the optical), $r_{\rm obj}$ is the radius of the comet (300 $\pm$ 30 m; \citealt{Li2017}), and $\varrho$ is the photometric aperture size in arcseconds (Table \ref{t3}). $\eta$ is the ratio of the mean optical scattering cross-section of the coma dust ($C_{\rm c}$) to the nucleus cross-section ($C_{\rm n}$). We assumed a spherical nucleus with radius of $r_{\rm obj}$, i.e., $C_{\rm n}$ = $\pi r_{\rm obj}^{\rm 2}$. The $C_{\rm c}$ can be computed in the same manner as in \citet{Luu1992}:
\begin{equation}
p_{\rm R}C_{\rm c} = 2.24\times10^{22}\pi~(r_{\rm H}\Delta)^2~10^{0.4(m_\odot -~m_{\rm R}(1, 1, 0))},
\label{eq:eq10}
\end{equation}
\noindent in which $p_{\rm R}$ is the geometric albedo (0.04 was assumed). As a result, the average $\dot{M}_{\rm d}$ prior to the first activation is 0.4 $\pm$ 0.2 kg s$^{\rm -1}$, increasing by $\sim$5.5 times during the first activation. Note that the above estimates of $\dot{M}_{\rm d}$, assuming the size of the most efficient scatterers of $\sim$1 $\mu$m in the optical wavelength, could become significantly higher if we consider the large-sized dust particles (e.g., $\bar{a}$ = 100 $\mu$m--1 mm). Despite the simplified assumptions we employed, it would be informative to monitor a long-term variation of the activity level of the comet. We summarized our photometry in Table \ref{t3}.

\begin{figure*}[!t]
\centering
\includegraphics[width=13.5cm]{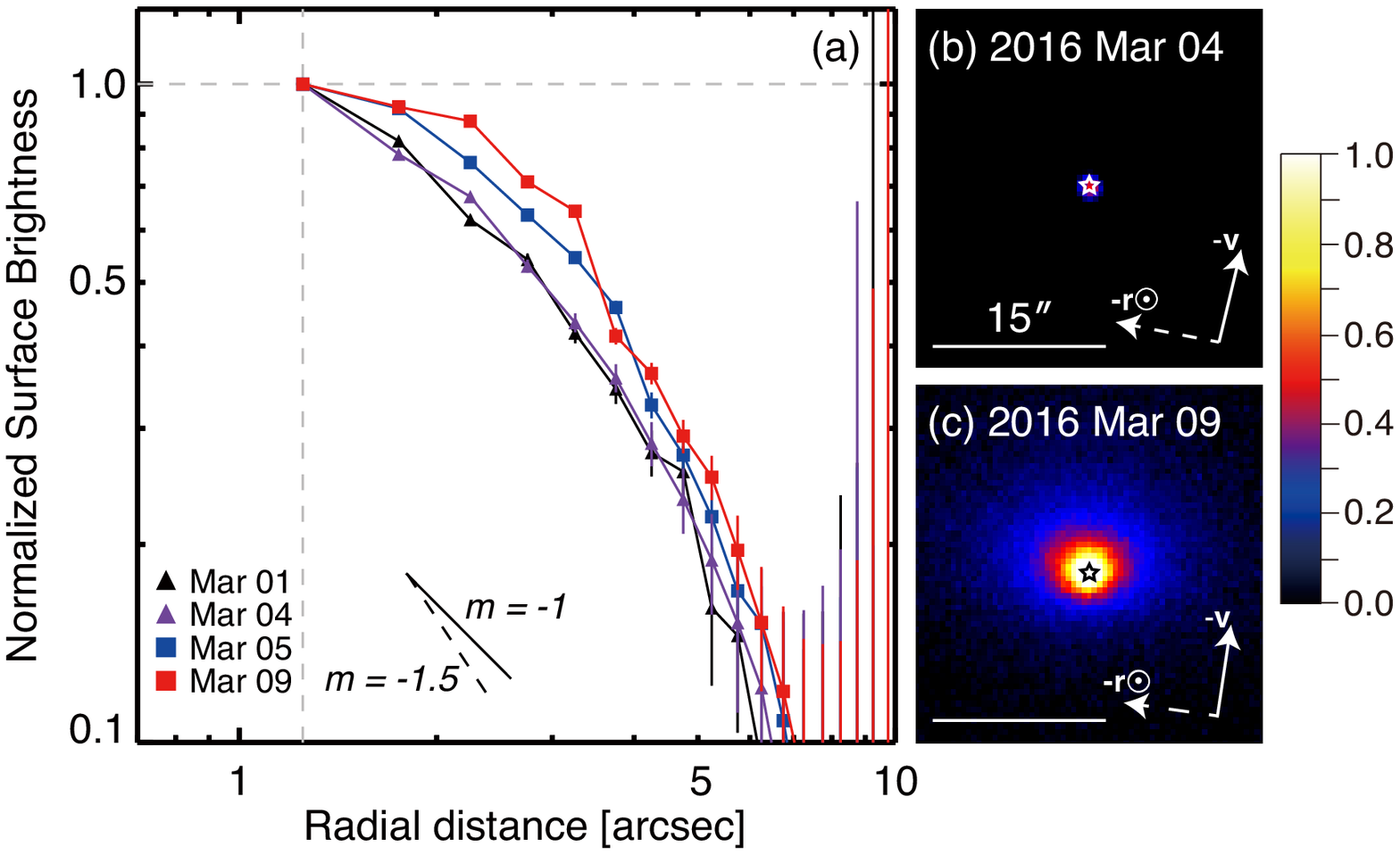}
\caption{Azimuthally averaged surface brightness profiles of 252P from the (a) LSGT ($R$ band) and IRSF ($H$ band; Stokes $I$) data. All brightness points are normalized to the radial distance of 1.25\arcsec. The black solid and dashed lines denote the gradients of $m$ = $-$1 and $m$ = $-$1.5, respectively, which are typical of cometary dust steadily expanding with initial ejection velocity under the solar radiation field \citep{Jewitt1987}. 2 $\times$ 2 binned $H$-band images are shown, each of which are from (b) before and (c) after the first activation of 252P taken from IRSF. Brightness was normalized between 0 and 1. North is up, and east is to the left. The bottom solid lines scale with 15$\arcsec$, and the dashed and solid arrows denote the antisolar ({\bf $-$r$_\odot$}) and negative velocity ({\bf $-$v}) vectors, respectively. Central stars mark the position of the nucleus.} 
\label{Fig2}
\end{figure*}

The above loose estimate of the activating point from the optical photometry (between March 01--05) would be narrowed using the Stokes $I$ maps of the IRSF data taken UT 2016 March 04 and 09 (open circles in Figure \ref{Fig1}a). Figure \ref{Fig2}a shows azimuthally averaged surface brightness profiles of 252P from the LSGT ($R$ band) and IRSF ($H$ band) data on a logarithmic scale.  All brightness points were normalized to the radial distance of 1.25\arcsec. Radial gradients of the points were then compared with the slopes of $m$ = $-$1 and $m$ = $-$1.5, which are typical of cometary dust expanding with initial ejection velocity under the solar radiation field \citep{Jewitt1987}. Compared to the triangles, both were decreasing steadily along the $m$ = $-$1 slope to $\sim$4\arcsec\ and along the $m$ = $-$1.5 slope outwards; squares therein show flatter distributions, and in particular, red squares on March 09 clearly show a shallower slope in the inner coma. 2 $\times$ 2 binned $H$-band images of the IRSF data are shown in panels (b) and (c), with the surface brightness being normalized between 0 and 1. A development of the central whitish-yellow part before versus after the activation is evident. On March 04, a feeble coma encircled the nucleus. The dust coma on March 09, however, was enlarged, being broadly elongated in the direction of the negative velocity vector ({\bf $-$v}) with respect to the nucleus position (star symbol).

Based on (i) the sudden increases of the photometric parameters of the LSGT data between March 01 and 05 and of the IRSF data between March 04 and 09 but (ii) nearly identical and steady radial profiles of the comet on March 01 and 04, we concluded that the first brightness enhancement most likely occurred on UT 2016 March 04--05. Similarly, we presented the multiband photometric results of the postperihelion reactivation of 252P that occurred on UT 2016 March 27--28 in the Appendix \ref{sec:postphot}. 

\subsection{Photometric results II: NIR dust color \label{sec:irsfphot}}

\begin{figure}[!b]
\centering
\includegraphics[width=9cm]{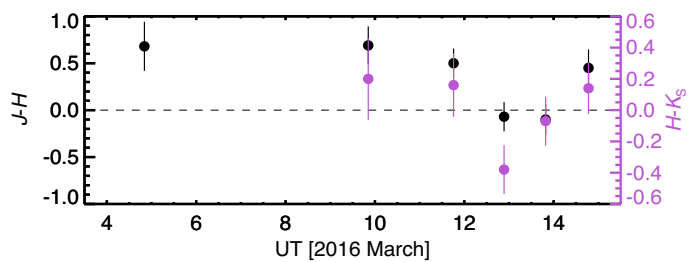}
\caption{Temporal evolution of the NIR color indices ($m_{\rm diff}$) of 252P dust measured from the IRSF data (Table \ref{t2}). Black and purple circles denote the colors of $J-H$ and $H-K_{\rm S}$, respectively.}
\label{Fig3}
\end{figure}

From the $I$ of the IRSF data (Eq. \ref{eq:eq1}), we obtained the colorimetric results of 252P dust. Figure \ref{Fig3} shows temporal evolution of the $m_{\rm IRSF}$ difference of $J-H$ and $H-K_{\rm S}$ as black and purple circles, respectively. Values of the magnitude differences calculated from $m_{\rm IRSF}$ in Table \ref{t2} are listed in Table \ref{t4}. Overall, the temporal evolution in the NIR dust color of 252P appears to be in opposition to that of the comet's brightness. Before the activation (March 04.84), the $J-H$ dust color was 0.68 $\pm$ 0.26 mag, which is consistent with the color range of cometary dust measured by \citet{Jewitt1986}. As time passed soon after the ignition, both $J-H$ and $H-K_{\rm S}$ values first decreased to the minima: from 0.69 $\pm$ 0.20 mag (March 09.85) and 0.50 $\pm$ 0.16 mag (March 11.76) to $-$0.07 $\pm$ 0.16 mag (March 12.89) and $-$0.10 $\pm$ 0.16 mag (March 13.82) for the $J-H$, and from 0.20 $\pm$ 0.19 mag (March 09.85) and 0.16 $\pm$ 0.22 mag (March 11.76) to $-$0.38 $\pm$ 0.16 mag (March 12.89) and $-$0.07 $\pm$ 0.15 mag (March 13.82) for the $H-K_{\rm S}$. The bluest color (i.e., minimal $m_{\rm IRSF}$ difference) occurred on March 12--13 both for the $J-H$ and $H-K_{\rm S}$, when the comet showed the maximum brightness (Figure \ref{Fig1}a). Subsequently, the colors reddened back to the pre-activation values: 0.45 $\pm$ 0.20 mag (March 14.78) and 0.47 $\pm$ 0.19 mag (March 15.83) for the $J-H$, and 0.14 $\pm$0.22 mag (March 14.78) and 0.12 $\pm$ 0.21 mag (March 15.83) for the $H-K_{\rm S}$.  

\begin{table*}[!t]
\centering
\caption{NIR photometric and polarimetric color indices of 252P/LINEAR from the IRSF data}
\begin{tabular}{c|c|cc|cc}
\toprule
Median UT & Epoch$^{\rm 1}$ & \multirow{2}{*}{$J-H$ $^{\rm 2}$ [mag]} & \multirow{2}{*}{$H-K_{\rm S}$ $^{\rm 3}$ [mag]} &  \multirow{2}{*}{{\it PC}$_{(J-H)}$$^{\rm 2}$ [\% $\mu$m$^{\rm -1}$]} & \multirow{2}{*}{{\it PC}$_{(H-K_{\rm S})}$$^{\rm 3}$ [\% $\mu$m$^{\rm -1}$]} \\
2016+ & number & & & & \\
\midrule
\midrule
03/04 20:16 & 1 & 0.68 $\pm$ 0.26 & $-$ & 0.13 $\pm$ 1.68 & $-$ \\
03/09 20:27 & 2 & 0.69 $\pm$ 0.20 & 0.20 $\pm$ 0.19 & $-$4.90 $\pm$ 1.34 & 0.03 $\pm$ 1.38 \\
03/11 18:12 & 3 & 0.56 $\pm$ 0.16 & 0.16 $\pm$ 0.22 & $-$0.75 $\pm$ 0.72 & $-$0.32 $\pm$ 2.38 \\
03/12 21:19 & 4 & $-$0.07 $\pm$ 0.16 & $-$0.38 $\pm$ 0.16 & $-$2.13 $\pm$ 1.69 & $-$7.10 $\pm$ 1.60 \\
03/13 19:43 & 5 & $-$0.10 $\pm$ 0.16 & $-$0.07 $\pm$ 0.15 & $-$2.15 $\pm$ 1.54 & $-$0.58 $\pm$ 1.51 \\
03/14 18:49 & 6 & 0.45 $\pm$ 0.20 & 0.14 $\pm$ 0.22 & $-$3.33 $\pm$ 1.42 & 0.27 $\pm$ 1.39 \\
03/15 19:56 & 7 & 0.47 $\pm$ 0.19 & 0.12 $\pm$ 0.21 & $-$2.05 $\pm$ 1.54 & $-$0.70 $\pm$ 1.49 \\
\bottomrule
\end{tabular}
\tablefoot{
\tablefoottext{\rm 1}{ Number of the epoch described in Figure \ref{Fig7};}
\tablefoottext{\rm 2}{ Magnitude difference measured between the $J$ and $H$ bands;}
\tablefoottext{\rm 3}{ Magnitude difference measured between the $H$ and $K_{\rm S}$ bands;}
\tablefoottext{\rm 4}{ Polarimetric color of the cometary dust over the $J$--$H$ bands;}
and \tablefoottext{\rm 5}{ Polarimetric color of the cometary dust over the $H$--$K_{\rm S}$ bands.}
}
\label{t4}
\vskip-1ex
\end{table*}

\subsection{Polarimetric results I: phase angle dependence \label{sec:pol1}}

Low albedos and porous aggregate structures of cometary dust particles have led to a general dependence of $P$ with respect to the scattering plane ($P_{\rm r}$) on the phase angle $\alpha$, parameterized by a shallow branch of negative polarization with an average minimum polarization $P_{\rm min}$ $\approx$ $-$1.5 \% at $\alpha_{\rm min}$ $\approx$ 10 \degr, an inversion angle $\alpha_{\rm 0}$ at $\alpha$ $\approx$ 22\degr, and a maximum polarization $P_{\rm max}$ $\approx$ 25--30 \% at $\alpha_{\rm max}$ $\approx$ 95\degr\ in the optical and NIR \citep{Kiselev2015}. 

\begin{figure}[!t]
\centering
\includegraphics[width=9cm]{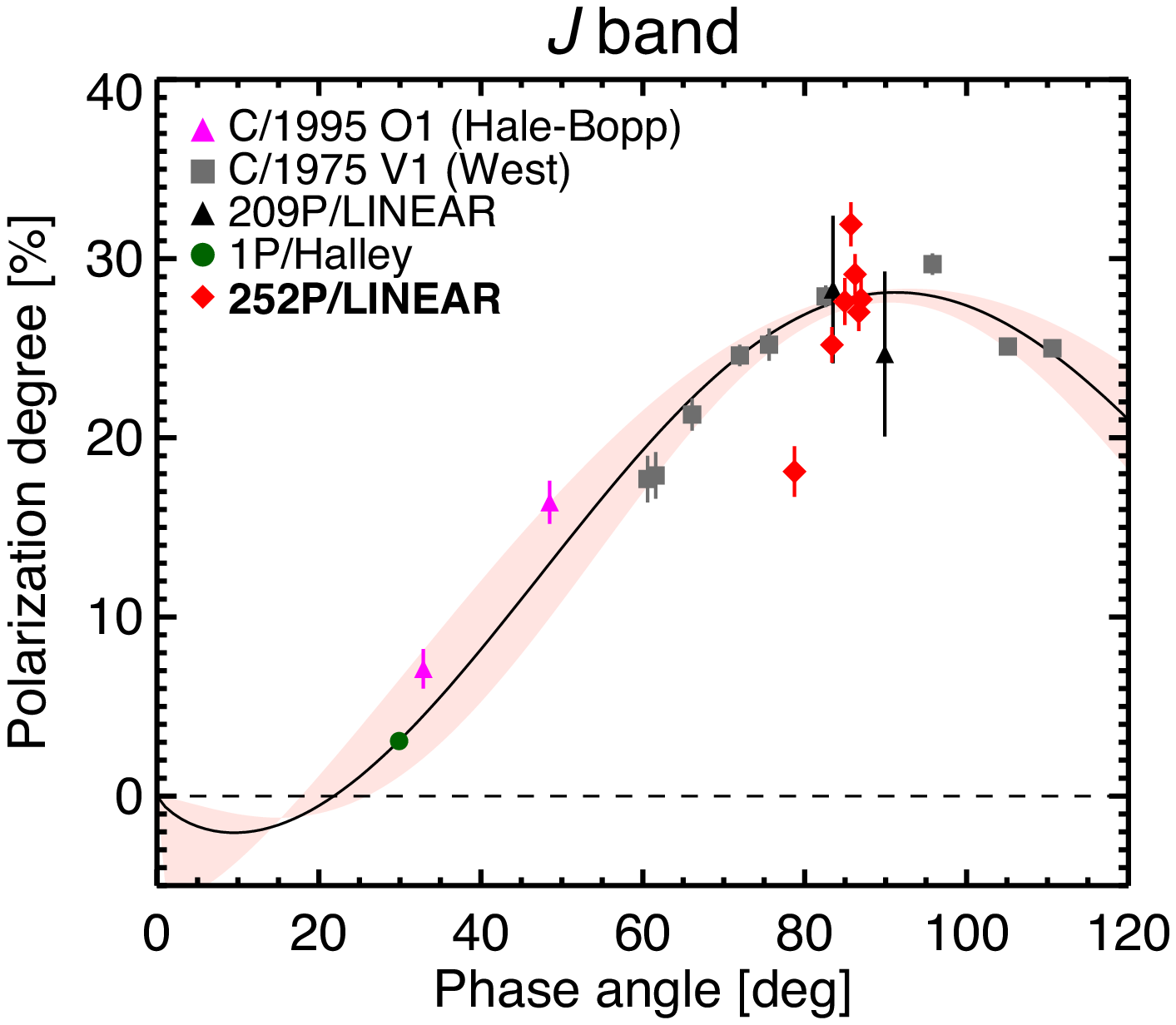}
\caption{$P_{\rm r}$ of comets in the $J$ band ($\lambda_{\rm C}$ = 1.25 $\mu$m) as a function of phase angle. The data for comets C/1995 O1 (Hale-Bopp), C/1975 V1 (West), and 1P/Halley are quoted from the database of comet polarimetry (DOCP; \citealt{Kiselev2010}), and the data for comets 209P/LINEAR and C/2013 US10 (Catalina) are from \citet{Kuroda2015} and \citet{Kwon2017}, respectively. The red symbols show the results for 252P. The solid curve denotes the interpolated average $\alpha$ dependence of the comets, as a result of Eq. \ref{eq:eq11}. The rose-colored area covers the 3$\sigma$ region of the average trend. The best fit (minimum $\chi$$^{\rm 2}$) parameters weighted by the square of the errors in $J$ band are $b$ = 38.64 $\pm$ 1.74 \%, $c_{\rm 1}$ = 0.77 $\pm$ 0.31, $c_{\rm 2}$ = 0.70 $\pm$ 0.11, and $\alpha_{\rm 0}$ = 21.85\degree $\pm$ 2.02\degree.} 
\label{Fig4}
\vskip-1ex
\end{figure}

\begin{figure}[!b]
\centering
\includegraphics[width=9cm]{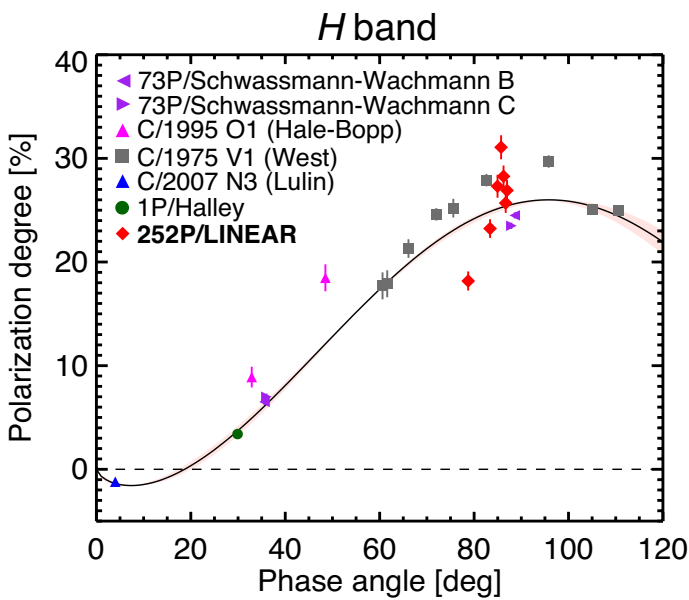}
\caption{Same as Figure \ref{Fig4} but in the $H$ band ($\lambda_{\rm C}$ = 1.65 $\mu$m). All the comet data except those for 252P are quoted from the DOCP. The best fit (minimum $\chi$$^{\rm 2}$) parameters weighted by the square of the errors in $H$ band are $b$ = 26.57 $\pm$ 0.16 \%, $c_{\rm 1}$ = 0.63 $\pm$ 0.04, $c_{\rm 2}$ = (3.41 $\pm$ 0.14) $\times$ 10$^{\rm -9}$, and $\alpha_{\rm 0}$ = $\alpha_{\rm 0}$ = 18.76\degree $\pm$ 0.41\degree.}
\label{Fig5}
\vskip-1ex
\end{figure}

\begin{figure}[!h]
\centering
\includegraphics[width=9cm]{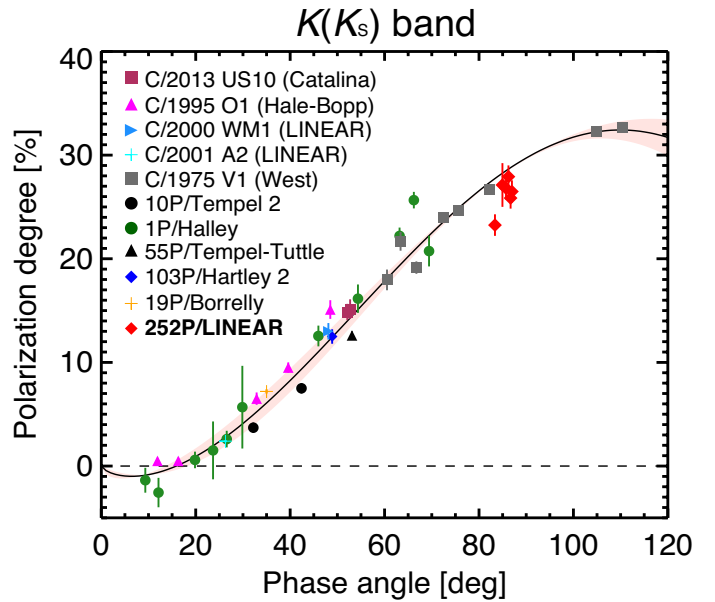}
\caption{Same as Figure \ref{Fig4} but in the $K$ and $K_{\rm S}$ bands ($\lambda_{\rm C}$ = 2.20 and 2.25 $\mu$m, respectively). All comet data with the exception of those for C/2013 US10 (Catalina) \citep{Kwon2017} and 252P are quoted from the DOCP. The best fit (minimum $\chi$$^{\rm 2}$) parameters weighted by the square of the errors in $K$ and $K_{\rm S}$ bands are $b$ = 31.47 $\pm$ 0.18 \%, $c_{\rm 1}$ = 0.65 $\pm$ 0.06, $c_{\rm 2}$ = (6.55 $\pm$ 0.18) $\times$ 10$^{\rm -10}$, and $\alpha_{\rm 0}$ = 16.97\degree $\pm$ 0.44\degree.}
\label{Fig6}
\vskip-1ex
\end{figure}

To investigate the polarimetric behavior of 252P, we first gleaned the archival NIR polarimetric data of cometary dust from the database of comet polarimetry (DOCP; \citealt{Kiselev2010}, as shown in Figure 22.3 of \citealt{Kiselev2015}) and later literature \citep{Kuroda2015,Kwon2017}. Fitting the average phase curve in each NIR band was obtained by employing the empirical trigonometric function of \citet{Penttila2005}:
\begin{equation}
P_{\rm r}(\alpha) = b~(\sin \alpha)^{c_{\rm 1}} \times \cos \left(\frac{\alpha}{2}\right)^{c_{\rm 2}} \times \sin (\alpha - \alpha_{\rm 0})~,
\label{eq:eq11}
\end{equation}
\noindent where $b$, $c_{\rm 1}$, $c_{\rm 2}$, and $\alpha_{\rm 0}$ are the wavelength-dependent parameters for characterizing the $P_{\rm r}$--$\alpha$ dependence. The best fit (minimum $\chi$$^{\rm 2}$) parameters weighted by the square of the errors in the $J$, $H$, and $K$ ($K_{\rm S}$) bands are described in the captions of Figures \ref{Fig4}, \ref{Fig5}, and \ref{Fig6}, respectively. We consider that the fitting results of the inversion angle and maximum polarization degrees are less reliable because of the few or lack of data observed at small ($\alpha$ $\lesssim$ 25\degr) and large ($\alpha$ $\gtrsim$ 110\degr) phase angle regions, despite the small fitting errors. Figures \ref{Fig4}--\ref{Fig6} present $P_{\rm r}$ of comets in the $J$, $H$, and $K$ (and $K_{\rm S}$) bands, respectively, as a function of phase angle $\alpha$. The average $\alpha$ dependencies are shown as the solid curves with colored 3$\sigma$ areas. The curves are basically the results of `interpolation'; thus, the error ranges of the four fitting parameters are not as large, except for the case of the $J$ band, in which there are no available data points at $\alpha$ < 30\degr. In this case, we forced the curve to fit the point (0, 0) just for visualization.


At first glance, the $P_{\rm r}$ values of cometary dust distribute quite homogeneously along the average phase curves, especially in the $K$ and $K_{\rm S}$ bands, regardless of the dynamical class of comets or the observational conditions (e.g., the observing geometry). The similarity in $P_{\rm r}$ may imply that dust grains of comets have similarities to a large extent in the physical (e.g., structure and size distribution) and/or compositional properties (e.g., complex refractive index). After a careful examination, however, we found that some comets still show deviations from the average trends. In particular, long period comet C/1995 O1 (Hale-Bopp) consistently shows higher $P_{\rm r}$($\alpha$) values than the majority of comets in all NIR bands \citep{Jones2000}, let alone in the optical \citep{Kwon2017}, whereas short period comets 10P/Tempel 2 and 55P/Tempel-Tuttle \citep{Kelley2004} apparently have slightly lower $P_{\rm r}$($\alpha$) values than the average in the $K$ band.

It is noteworthy that 252P showed an abrupt change in $P_{\rm r}$($\alpha$) during our observations, especially before and soon after the sudden activation on March 04--05. In Figures \ref{Fig4} and \ref{Fig5}, $P_{\rm r}$ values on March 04 (i.e., before the first activation) were significantly lower (by $\sim$7 \% in the $J$ and by $\sim$5 \% in the $H$ bands, but no available $K_{\rm S}$ band data due to the low SNR of < 3) than the expected $P_{\rm r}$ at given $\alpha$ (i.e., the fitted curves). Throughout the activation, however, $P_{\rm r}$ increased by $\sim$13 \% to the maxima on March 12--13. The $P_{\rm r}$ at perihelion ($\alpha$ $\approx$ 87\degr\ on March 15.38) seems to return to normal. Such a temporal $P_{\rm r}$($\alpha$) change of the comet apparently looks similar to the patterns of photometrically driven parameters, i.e., $m_{\rm R}$, Af$\rho$, and $\dot{M}_{\rm d}$(Section \ref{sec:phot1}). 

The polarization vectors of the comet ($\theta_{\rm r}$) were roughly aligned perpendicular to the scattering plane (within 4.6\degree$-$10.9\degree from the normal direction to the scattering plane) in the $J$, $H$, and $K_{\rm S}$ bands but presented a decreasing trend as $\alpha$ increased ($\theta_{\rm r}$ in Table \ref{t2}). It should be expected that a randomly distributed scattering medium tends to signal its strongest intensity of the polarized light to the normal direction of the scattering plane (i.e., $\theta_{\rm r}$ $\sim$ 0\degree; \citealt{Bohren1983}). Breaking such a condition of coma dust might lead to a nonzero value and some systematic trends of $\theta_{\rm r}$; however, we could not provide a conclusive explanation for this phenomenon from our results. It might be due to an increase in the accuracy of measurements as the values of the Stokes parameters increase. Alternatively, one of the causes might come from a substructure in the coma of 252P, indecomposable by the aperture polarimetry in this study. A series of high-resolution images taken by HST on March 14 and April 04 2016 suggested that 252P had a strong sunward jet, the influence of which vanished at an aperture size of >50--60 pixels (corresponding to 80--100 km in radial distance). As mentioned in Section \ref{sec:data}, our aperture size integrated all signals within the projected radial distances of 82 km (on March 04) to 173 km (on March 15) during the observations, which is similar to or slightly larger than the jet scales. 

\begin{figure}[!b]
\centering
\includegraphics[width=7cm]{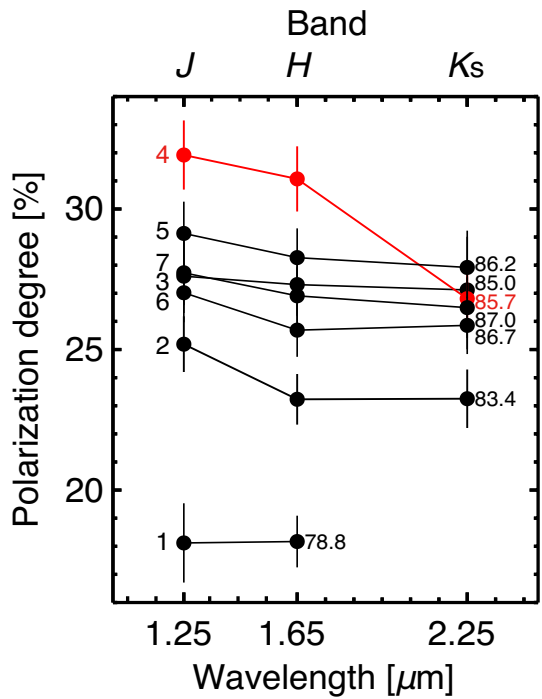}
\caption{$P_{\rm r}$ of 252P from the IRSF data (Table \ref{t2}) as a function of the wavelength. Numbers on the left side of the data indicate the order of the observation epochs (1: March 04, 2: March 09, 3: March 11, 4: March 12, 5: March 13, 6: March 14, and 7: March 15), and numbers on the right side are the phase angles of the corresponding epochs. Red on the fourth epoch (March 12) denotes the day when the comet shows the maximum NIR brightness (open circles in Figure \ref{Fig1}a). }
\label{Fig7}
\vskip-1ex
\end{figure}

\subsection{Polarimetric result II: spectral dependence \label{sec:pol2}}
In addition to the $\alpha$-dependence, $P_{\rm r}$ exhibits spectral dependence (so-called `polarimetric color', {\it PC}), which is defined as
\begin{equation}
PC = \frac{\Delta P_{\rm r}}{\Delta \lambda} = \frac{P_{\rm r}(\lambda_{\rm 2}) - P_{\rm r}(\lambda_{\rm 1})}{\lambda_{\rm 2} - \lambda_{\rm 1}},
\label{eq:eq12}
\end{equation}
\noindent where $P_{\rm r}$($\lambda_{\rm 1}$) and $P_{\rm r}$($\lambda_{\rm 2}$) are the $P_{\rm r}$ values in percent measured at the center wavelengths of $\lambda_{\rm 1}$ $\mu$m and $\lambda_{\rm 2}$ $\mu$m, respectively, and $\Delta \lambda$ is the difference between the two wavelengths ($\lambda_{\rm 2}$$-$$\lambda_{\rm 1}$, when $\lambda_{\rm 2}$ > $\lambda_{\rm 1}$) in $\mu$m. Positive {\it PC} is conventionally labeled as `red', and negative {\it PC} is labeled as `blue'. {\it PC} also depends on the phase angle, but, in general, {\it PC} of cometary dust is red over $\lambda$ = 0.5--1.6 $\mu$m and seems to turn blue at longer wavelengths at $\alpha$ > 25\degr\ \citep{Kolokolova2004}.

Figure \ref{Fig7} shows the polarimetric color of 252P from the IRSF data. A sequence of the observation epochs and corresponding phase angles are shown on the left and right sides of the data points, respectively. Red on the fourth epoch (March 12) indicates the day when 252P showed the maximum NIR brightness during the IRSF observation (Figure \ref{Fig1}a). The derived values of the {\it PC} and their errors are listed in Table \ref{t4}. Before the activation, {\it PC} was nearly neutral (1; March 04). After the activation (2--7; March 09--15), however, blue {\it PC} became dominant over the $J$ and $H$ bands (i.e., $-$2.55 \% $\mu$m$^{\rm -1}$ on average), while neutral-to-blue {\it PC} was shown over the $H$ and $K_{\rm S}$ bands (i.e., $-$0.26 \% $\mu$m$^{\rm -1}$ on average), except for the steep negative slope of {\it PC} at the fourth epoch (4; red line on March 12).

Blue-dominating {\it PC} of 252P over the $J$ and $H$ bands is distinctively different from other comets observed at high phase angles, which showed moderate $P_{\rm r}$ increase (e.g., $\sim$2.9 \% $\mu$m$^{\rm -1}$ for comet 1P/Halley at $\alpha$ = 65 \degr and $\sim$3.1 \% $\mu$m$^{\rm -1}$ for comet C/1975 V1 (West) at $\alpha$ = 66\degr; \citealt{Kiselev2015}) at this domain. Overall, 252P exhibits similar blue {\it PC} trends at NIR during the activation, except for the steepest negative {\it PC} over the $H$ and $K_{\rm S}$ bands at the fourth epoch (red points) when the comet exhibited the maximum brightness with the highest $P_{\rm r}$ in shorter wavelengths (open circles in Figure \ref{Fig1}a). Plausible scenarios to produce the observed polarimetric properties of the comet, together with the colorimetric behaviors, will be discussed in Section \ref{sec:dis1}.

\section{Discussion \label{sec:discuss}}

\subsection{Abrupt change in polarization degrees near the first activation \label{sec:dis1}}
As shown in Sections \ref{sec:pol1} and \ref{sec:pol2}, both NIR polarization degree $P_{\rm r}$ and polarimetric color {\it PC} of 252P precipitously changed upon the first activation point only within the change in $\delta$$r_{\rm H}$ = 0.008 au, which is one of the biggest and rapid changes ever observed. It is reminiscent of comet D/1999 S4 (LINEAR), a totally disintegrated comet whose $P_{\rm r}$ in the optical increased sharply around its perihelion \citep{Kiselev2002}. By definition, $P_{\rm r}$ is the ratio of the difference of light intensities measured between perpendicular and parallel directions against the scattering plane to the total intensity. A change in $P_{\rm r}$ thus indicates a change of physical and/or compositional properties of dust particles responsible for light scattering as the comet approached the Sun. Since we observed the comet with the identical instrument in simultaneous multiband imaging polarimetric mode, therewith correcting the systematic effects, it is likely that we have detected the real phenomena. We considered three cases that could be affecting the polarimetric variations, together with the NIR colorimetric results (Section \ref{sec:irsfphot}): changes in the (i) composition, (ii) effective size and (iii) porosity of dust particles. 

\subsubsection{Change in composition of dust particles \label{sec:dis1-1}}
The primary constituents of cometary dust are silicates (mainly in the form of olivine and pyroxene), carbonaceous materials (amorphous carbon and organics), Fe-bearing sulfides, and ice \citep{Levasseur-Regourd2018}. Unlike asteroids, the optical properties of which are largely differentiated with respect to heliocentric distance \citep{DeMeo2014,Belskaya2017}, comets seem to possess fairly homogeneous bulk optical properties with a low geometric albedo of $\sim$0.04 \citep{Lamy2004}, showing a nearly uniform distribution on the polarimetric phase curves, regardless of dynamical classes \citep{Kwon2017}. However, any difference in a comet might be expected if we consider the grain properties with depth in the nucleus, perhaps as a result of mainly solar irradiation on the outer layers. Notably, solar heating tends to deplete near-surface volatiles (e.g., \citealt{Prialnik2004}). As such, the observed discontinuous brightening of 252P (Section \ref{sec:phot1}), as well as its jet structure detected in a high-resolution images \citep{Li2017} in 2016 apparition, might cast a possibility for fresh particles to be ejected from the interior of the nucleus.
 
Laboratory experiments and numerical modeling of comet analogs showed that changes in dust compositions could make a difference to total absorptivity, which leads to a change in polarimetric properties: more transparent particles are subject to multiple scattering, so that the scattered light would be more depolarized. Absorptive particles, however, suppress the multiple scattering, which results in higher $P_{\rm r}$. Hence, increasing absorption with wavelength may lead to red {\it PC}, whereas decreasing absorption with wavelength may cause blue {\it PC} while all other properties (i.e., size and porosity) remained unchanged (\citealt{Gustafson1999,Kolokolova2001,Kolokolova2004,Kimura2006} and references therein). 

Therefore, to satisfy the observed increase in $P_{\rm r}$ of 252P, first, the ejection of more absorptive particles is required. For carbonaceous materials, their absorption coefficients on wavelength increases in the NIR \citep{Rouleau1991,Greenberg1996,Kolokolova1997}, such that they produce a red {\it PC}, not our observed blue {\it PC}. On the contrary, the ejection of the compact icy or solid absorbing particles \citep{Warren1984,Warren2019}, whose scattering largely depends on the single scattering of the whole particles but not as much on the absorption coefficients as fluffy ones \citep{Kolokolova1997,Gustafson1999}, makes them contribute to the increase in $P_{\rm r}$ and probably to the blue {\it PC} if the size of the particles are as large as $>$a few tens of $\mu$m (see Section \ref{sec:dis1-2} for detail discussions on the size and porosity). The idea of the ejection of compact icy particles during the discontinuous brightening around the perihelion might be reconciled with the existence of sintered subsurface icy layer \citep{Kossacki1994,Kossacki2015}. Meanwhile, the bluing of the NIR color data (Section \ref{sec:irsfphot}) could put an additional constraint, if this is a purely compositional effect, that more transparent materials were ejected during the activation. However, the dominance of such material is incompatible with the sharp increase of $P_{\rm r}$, which indicates that factors other than the compositional effect should be the primary cause of the NIR color agent.

In summary, variations in the composition of dust particles may not be a primary factor provoking the sudden change in polarimetric parameters of 252P preperihelion, but instead other properties, such as porosity and size of dust particles, should be considered as the responsible factors for the observed phenomena of the comet.

\subsubsection{Change in particle size and porosity \label{sec:dis1-2}}
Scattering regimes can be broadly classified into three types depending on the size parameter $X$ of dust ($X$ := 2$\pi$$a$ / $\lambda$, where $a$ is the particle radius and $\lambda$ is the observation wavelength): the Rayleigh, Mie, and geometrical optics regimes in order of increasing $a$. Relatively high $P_{\rm r}(\alpha)$ can be achieved in both the Rayleigh ($X$ $\ll$ 1; $a$ $\ll$ 0.2 $\mu$m in the $J$, $a$ $\ll$ 0.3 $\mu$m in the $H$, and $a$ $\ll$ 0.4 $\mu$m in the $K_{\rm S}$ bands) and geometrical optics ($X$ $\gg$ 10; $a$ $\gg$ 2.0 $\mu$m in the $J$, $a$ $\gg$ 2.6 $\mu$m in the $H$, and $a$ $\gg$ 3.6 $\mu$m in the $K_{\rm S}$ bands) regimes. 

For Rayleigh-like tiny particles in the Mie scattering regime ($X$ $\sim$ 1--10; an order of 0.1--1 $\mu$m dust in the NIR), the greater their contribution to the scattered light is, the higher the observed $P_{\rm r}(\alpha)$ and the redder {\it PC} \citep{Kolokolova2004}. Unfortunately, the dominance of such particles would not explain the observed results for 252P, particularly the observed bluing of the NIR dust color (Figure \ref{Fig3}) and prevailing blue {\it PC} over the $J$ and $H$ bands upon the activation (Figure \ref{Fig7}), and seems to conflict with the results of previous studies showing a minor role of very small-sized particles in mass density, i.e., dynamics \citep{Fulle2015,Rotundi2015,Fulle2016}, and in light scattering of cometary dust \citep{Jewitt1986,Kolokolova2007}. Therefore, we dismiss the possibility that the observed polarimetric changes were {\it largely} stimulated by the increase of Rayleigh(-like) particles, although not ruling out their possible existence evidenced by the change in the optical brightness (Section \ref{sec:phot1}) and observed jet-like structure in the optical \citep{Li2017}. Meanwhile, such small dust particles may thermally depolarize the $K_{\rm S}$ band data (e.g., blue {\it PC} over the $H$ and $K_{\rm S}$ bands in Figure \ref{Fig7}), but we precluded the predominance of such a possibility based on (i) the $P_{\rm r}(\alpha)$ of 252P in the $K_{\rm S}$ band near the perihelion having similar values to the points of comet C/1975 V1 (West) (gray squares; \citealt{Oishi1978}) from which thermal flux was subtracted and on (ii) the case of C/1975 V1 (West), the thermal depolarization effect of which was negligible at $r_{\rm H}$ $\gtrsim$ 0.9 au \citep{Oishi1978}.

Large dust particles in the geometrical optics regime ($a$ > a few tens of $\mu$m) can be further considered with two different porosities: fluffy (porous) and compact particles, two of which are in fact two main populations of dust particles of comet 67P/Churyumov-Gerasimenko (67P) collected by Rosetta/GIADA \citep{Fulle2015}. In case of large fluffy particles (tensile strength of <10$^{\rm 5}$ N m$^{\rm -2}$ (\citealt{Mendis1991}) and high charge-to-mass ratio (\citealt{Fulle2015})), they behave similarly to that of the individual constituent particles in dynamics \citep{Mukai1992}, in light scattering \citep{Kolokolova2011}, and in thermal emissions \citep{Wooden2002,Kolokolova2007}. Accordingly, the polarimetric properties of scattered light by such large fluffy dust particles would be akin to those by particles approaching the Rayleigh regime (but are larger than $X$ = 1) (i.e., an increase in $P_{\rm r}$ with the red {\it PC}), which again contradicts the observed results of 252P.

Finally, we consider to what extent large, compact dust particles contributed to the observed polarimetric properties of 252P. A dominance of such particles was corroborated by the results of Rosetta/GIADA, showing that mm-to-cm-scaled compact chunks ejected near perihelion, along with the release of the upper desiccated layer with a higher refractory-to-ice mass ratio than that inside the nucleus, accounted for $>$85 \% of the dust mass loss and luminosity function of the comet (e.g., \citealt{Blum2017,Fulle2018}). As the interaction energy between ambient dipoles (i.e., monomers) are inversely proportional to the cube of the distance between the two \citep{Jackson1965}, less porous materials are more subject to the scattered light from neighboring dust constituents. Accordingly, as the wavelength of the incident light increases (i.e., more monomers are covered in a single wavelength), $P_{\rm r}$ of such compact particles decreases more rapidly than that of the porous particles due to the enhanced electromagnetic interaction \citep{Gustafson1999,Kolokolova2010}. Compared to the non/long-periodic comets observed at similar phase angles, it is more likely that the near-surface of 252P has a paucity of small, fluffy dust particles owing to its more frequent perihelion passages in the near-Earth orbit (e.g., \citealt{Li1998,Kolokolova2007}). The circumstance would emerge as the unusual blue {\it PC} of the comet. For this reason, the dominance of signals by the ejection of large, compact particles into the coma would be attributable to both a sudden increase in $P_{\rm r}$ and the enhanced blue {\it PC} upon the first activation of 252P. 

\subsubsection{Tentative conclusions on particle properties ejected during the activation}
The ejection of large (i.e., $a$ is at least >a few tens of $\mu$m in the geometrical optics regime) and compact (likely as low as the porosity of $\sim$30--65 \% of the near-surface Rosetta/Philae landing site of 67P; \citealt{Spohn2015}) particles can also be deduced from the morphology of the dust coma and the color change of the apparent magnitudes of the IRSF data. Figure \ref{Fig2}c shows a broad extension of the dust cloud to the trailing direction ({\bf -v}) with respect to the photocenter, alluding to the ejection of dust particles that are insensitive to the solar radiation pressure at $r_{\rm H}$ $\lesssim$ 1 au. As such, a primary factor to distribute the ejected particles in the {\bf -v} direction should rather be an explosive mechanism, such as a rocket force from sublimating gas with nonzero ejection velocity \citep{Kelley2013}, which is most likely responsible for the discontinuous brightening of the comet. Concurrently, the $J-H$ and $H-K_{\rm S}$ colors of the comet decreased with time (Figure \ref{Fig3}): from the typical color of cometary dust (e.g., \citealt{Jewitt1986}) before the activation to the blue--neutral color. It may also support our scenario that the dust particles, which are brighter at the longer wavelengths, contributed more to the intensity of the comet than the smaller or large porous dust particles. Taken as a whole, the above conjectures may answer our first question: the evolved near-surface dust layer would primarily define the activity of the comet as long as the incoming solar heat flux is sufficiently large.

\subsection{Implication for evolved surface \label{sec:dis2}}

Our second question is how useful NIR polarimetry would be to discriminate the behaviors of fresh and evolved dust particles. The discussion on the observed polarimetric properties largely induced by the evolved dust particles of 252P in Section \ref{sec:dis1} could advocate for the potential usefulness of this approach. 

Returning to Figure \ref{Fig6}, the $P_{\rm r}$($\alpha$) of 11 comets in the K band, where the number of observed comets to date is the largest among NIR bands, follows the average trend quite well, while a careful check reveals that non/long-periodic comets tend to show, on average, higher $P_{\rm r}$($\alpha$) values, particularly than for two Jupiter-Family comets (JFCs), 10P/Tempel 2 and 55P/Tempel-Tuttle \citep{Kelley2004}. Interestingly, the two are characterized by weak-to-absent 10 $\mu$m silicate emission features and subtle temperature excess for the blackbody temperature \citep{Lynch1995,Lynch2000}, both suggesting a lack of small $\mu$m-sized and/or fluffy dust particles \citep{Hanner1999,Wooden2002,Lisse2002}. Given that JFCs orbit closer to the Sun than non/long-periodic comets, which offers a more favorable environment to develop the consolidated dust mantles on their surfaces, it would be understandable for JFCs to show lower $P_{\rm r}$($\alpha$) at NIR than those of less-heated comets. The orbital evolution of 252P in the near-Earth orbit over 300 yr or even more would be in favor of this scenario (Appendix \ref{sec:orbit}).
However, for the moment, we should be cautious in drawing any firm conclusions. A single $P_{\rm r}$($\alpha$) point of a comet does not provide much information on its physical and compositional properties, but rather, we need (i) (quasi-)simultaneous multiband polarimetric data to measure the {\it PC} of dust particles for estimation of the porosity of dust and/or (ii) comets observed at multiple observation epochs, including the high $\alpha$ region, to trace the variation of the $P_{\rm r}$($\alpha$) and {\it PC}. Further NIR polarimetric observations undertaken in a well-organized manner are highly desirable to investigate secular evolutions of polarimetric parameters of cometary dust on the orbital motion as well as to search for any systematic differences in them between different dynamical groups of comets.

\section{Summary \label{sec:sum}}
We present multiband NIR imaging polarimetric observations around the perihelion passage and optical imaging observations of comet 252P/LINEAR taken over four months in its 2016 apparition. We summarized the main results as follows.

\begin{enumerate}

\item  We detected two discontinuous brightness enhancements of 252P: one in the inbound (on UT 2016 March 04--05) orbit and the other in the outbound (March 27--28) orbit. A month prior to the perihelion passage, 252P already showed $\sim$13 times higher $Af\rho$ values than that of the 2000 apparition. 

\item Upon the first activation, $m_{\rm R}(1, 1, 0)$ and $Af\rho$ of 252P derived from the optical $R$ and $R_{\rm C}$ bands data increased by $\sim$2 mag and $\sim$35 cm, respectively. The attendant $\dot{M}_{\rm d}$ for assumed 1 $\mu$m particles increased to $\sim$5.5 times the initial state. In the meantime, both the $J-H$ and $H-K_{\rm S}$ dust color of 252P had decreased, showing the bluest color in the middle of the activation.

\item  Before the first activation, the $P_{\rm r}$ values of 252P were far lower than the average trend of comets at given $\alpha$, by $\sim$7 \% and $\sim$5 \% in the $J$ and $H$ bands, respectively. Upon and soon after the activation, however, the $P_{\rm r}$ increased in all NIR bands by $\sim$13 \% at most, showing the distinctive development of the blue {\it PC} over 1.25--2.25 $\mu$m similar to, but fiercer than the behaviors of a fragmenting comet D/1999 S4 (LINEAR) \citep{Kiselev2002}. In particular, the blue-dominating {\it PC} at the $J$--$H$ bands ($-$2.55 \% $\mu$m$^{\rm -1}$ on average) is different from other comets observed, which show moderately red {\it PC} at similar phase angles. 

\item The most likely implication of the sudden change in the observed polarimetric properties of the comet during the activation (i.e., increase $P_{\rm r}$ with the blue {\it PC}) as well as the bluing of the NIR dust color would be the dominance of large (i.e., well located in the geometrical optics regime at NIR), compact particles predominantly ejected from the desiccated dust layer. The paucity of small, fluffy dust particles around the nucleus of 252P would be ascribed to the more intense solar heating effects the comet has received in the near-Earth orbit (Appendix \ref{sec:orbit}) than for the less-heated non/long-periodic comets observed to date.

\end{enumerate}

\begin{acknowledgements}
We thank the referee, L., Kolokolova, whose careful reading and constructive comments improved our manuscript. Y.G.K. was supported by the Global Ph.D. Fellowship Program through a National Research Foundation of Korea (NRF) grant funded by the Ministry of Education (NRF-2015H1A2A1034260). This work at Seoul National University was also supported by the NRF funded by the Korean Government (MEST), No. 2015R1D1A1A01060025. J.K. was supported by Astrobiology Center of NINS. M.T. was supported by MEXT/JSPS KAKENHI grant Nos. 18H05442, 15H02063, and 22000005. The IRSF project is a collaboration between Nagoya University and the South African Astronomical Observatory (SAAO) supported by Grants-in-Aid for Scientific Research on Priority Areas (A; No. 10147207 and No. 10147214) and Optical \& Near-Infrared Astronomy Inter-University Cooperation Program, from the Ministry of Education, Culture, Sports, Science and Technology (MEXT) of Japan and the National Research Foundation (NRF) of South Africa.
\end{acknowledgements}


\begin{appendix}
\section{Postperihelion activation of 252P \label{sec:postphot}}

\subsection{photometric results}
Changes in the postperihelion activity level of 252P were quantified by the photometry, as in Section \ref{sec:phot1}. The intrinsic brightness of 252P (Figure \ref{Fig1}a; derived from Eq. \ref{eq:eq7}) soon after perihelion passage seemed to decline steeply, but it began to rebound from the minimum of $\sim$17.3 mag with a nearly 16 times flux increase on UT 2016 March 27--28 (right arrow in Figure \ref{Fig1}a). Such re-ignition again became tranquil after peaking in approximately late April to early May at $\sim$14.3 mag. The changes in $Af\rho$ and $\dot{M}_{\rm d}$ patterns also showed similar trends with that of $m_{\rm R}$(1, 1, 0), despite the former two having their maxima $\sim$ a week after the peak of $m_{\rm R}$(1, 1, 0). 

\begin{figure*}[!t]
\centering
\includegraphics[width=12cm]{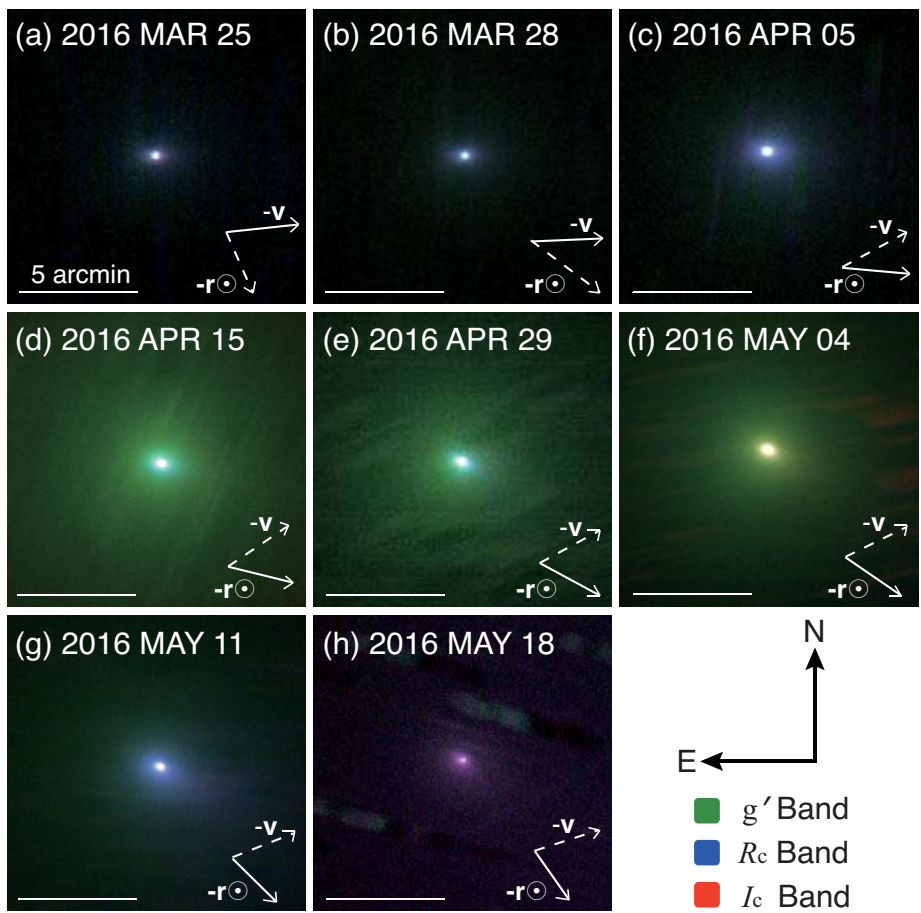}
\caption{Wide-field composite images of 252P from the multiband OAO data. In each panel, the observation date and 5$\arcmin$ scale bar are given at the top and bottom left, respectively. Denotations of arrows are identical with those in Figure \ref{Fig2}. North is up, and east is to the left. Green, blue, and red hues represent the fluxes in the g$'$, $R_{\rm C}$, and $I_{\rm C}$ bands, respectively. Note that the applied photometric aperture size of 1000 km in radius during this period decreased from 33.6\arcsec to 2.5\arcsec, which is much smaller than the field of view of each panel.}
\label{Fig9}
\vskip-1ex
\end{figure*}

Over 1.5 months after the second activation, $Af\rho$ of 252P (Figure \ref{Fig1}b; derived from Eq. \ref{eq:eq8}) increased by $\sim$57 times from 14.1 $\pm$ 3.8 cm to 808.7 $\pm$ 17.6 cm. We compared the $Af\rho$ values with the {\it Hubble Space Telescope} (HST) data in the broadband F625W filter ($\lambda_{\rm C}$ = 0.625 $\mu$m) taken on UT 2016 March 14 and April 04 (16.8 $\pm$ 0.3 cm and 57 $\pm$ 1 cm, respectively, from \citealt{Li2017}; cross points in Figure \ref{Fig1}b). The consistent differences by $\sim$20 cm between the HST and OAO data observed at similar epochs exist, but we suspect that the different photometric aperture sizes employed (by more than two orders of magnitude, i.e., $\rho$ $\lesssim$ 10 km for the HST data and $\rho$ = 1000 km for the OAO data) might cause the difference in $Af\rho$. Although, by definition, $Af\rho$ should be a constant value regardless of the aperture size under the steady expansion in the coma, such a condition would not be met in the vicinity of the nucleus. Considering that $Af\rho$ increases outward to reach an equilibrium (to $\sim$1000 km at $r_{\rm H}$ $\sim$ 1 au; see, e.g., \citealt{Bonev2008}), it might be explainable that the OAO data in this study consistently showed higher $Af\rho$ than those of the HST data. Nonetheless, the HST data can be interpreted in line with the OAO data, in that $Af\rho$ of 252P in postperihelion was higher than that in preperihelion.

Similarly, variations in $\dot{M}_{\rm d}$ (Figure \ref{Fig1}c; derived from Eq. \ref{eq:eq9} and Eq. \ref{eq:eq10}) strongly support the postperihelion reactivation of 252P, although the overall morphology of the evolutionary trend follows a concave shape, increasing and decreasing both in proportion to the third power of $r_{\rm H}$, unlike the convex shapes of the $m_{\rm R}$(1, 1, 0) and $Af\rho$ trends. From the minimum of 1.1 $\pm$ 1.0 kg s$^{\rm -1}$ on March 27 through the maximum of 31.3 $\pm$ 2.1 kg s$^{\rm -1}$ on May 02 to another minimum of 4.5 $\pm$ 2.4 kg s$^{\rm -1}$ on June 10, the total mass-loss during the second activation event was $\sim$7.8 $\times$ 10$^{\rm 7}$ kg, when we assumed nominal $\bar{a}$ = 1 $\mu$m dust grains. Taken as a whole, the derived parameters indicate that the second activation of 252P was likely on UT 2016 March 27--28.

\subsection{Variation in the coma color}
\begin{figure}[!b]
\centering
\includegraphics[width=9cm]{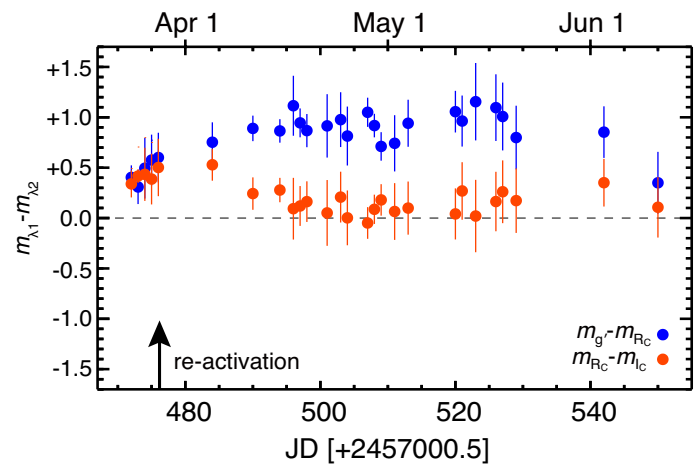}
\caption{Temporal evolution in the postperihelion absolute magnitudes of 252P from the multiband OAO data. $m_{\rm X}$ denotes the absolute magnitude in X band, with an aperture size of 1000 km (corresponding to  33.6\arcsec--2.5\arcsec) in radius. Differences in $m_{\rm X}$ between two bands are shown as blue ($m_{\rm g'}$ $-$ $m_{\rm R_{\rm C}}$) and orange ($m_{\rm R_{\rm C}}$ $-$ $m_{\rm I_{\rm C}}$) circles. All the error bars include the 1$\sigma$ errors on the measurements, calibration of the field stars with the assumed internal catalog error of $\sim$0.1, and propagated errors in magnitude subtraction of two bands.}
\label{Fig10}
\vskip-1ex
\end{figure}

As a byproduct of our multiband imaging observation, we produced color images to clarify to what extent the fluxes from each band contributed to the postperihelion reactivation of the comet. The g$'$, $R_{\rm C}$, and $I_{\rm C}$ band images were allocated as green, blue, and red hues, respectively. Since we employed the broadband filters, gas contaminations in flux estimation would be inevitable, particularly by the molecular emissions of C$_{\rm 2}$ Swan bands in the g$'$, NH$_{\rm 2}$ $\alpha$ bands in the $R_{\rm C}$, and CN red bands in the $I_{\rm C}$ filters \citep{Meech2004}. 

Figure \ref{Fig9} presents wide-field composite images, together with the dates of observation and 5$\arcmin$ scale bars in each panel, the fields of view of which are much larger than the applied photometric aperture size. In these images, greenish and bluish components spherically expanded from the nucleus once the comet became activated, whereas a whitish component elongated along the antisolar direction ({\bf $-$r$_\odot$}). As the activity of the comet peaked, a greenish coma covered the entire field of view, developing a significant bluish green color in the innermost coma. The gas coma of 252P at this term spherically extended to >10$^{\rm 4}$ km from the nucleus in the sky plane (panels (d) to (f)). Meanwhile, a reddish component was observably dominant, only after the reactivating event was over (panel (h)).

Figure \ref{Fig10} shows the temporal evolution of the 252P coma color in postperihelion more quantitatively. All reduced magnitudes were derived with the 1000 km aperture size (corresponding to 33.6\arcsec--2.5\arcsec) in radius, well inside the central-most regions of the images in Figure \ref{Fig9}. The magnitude differences between the g$'-R_{\rm C}$ and $R_{\rm C}-I_{\rm C}$ bands are given as blue ($m_{\rm g'}$ $-$ $m_{\rm R_{\rm C}}$) and orange ($m_{\rm R_{\rm C}}$ $-$ $m_{\rm I_{\rm C}}$) circles, respectively. The putative activation point (on March 27--28) is marked as an arrow. Overall, a blueward color change (i.e., predominant brightening in the $R_{\rm C}$ band flux) is apparent in the inner coma. Before the activation, the comet was slightly red, showing on average $m_{\rm g'}$ $-$ $m_{\rm R_{\rm C}}$ $\sim$ 0.5 and $m_{\rm R_{\rm C}}$ $-$ $m_{\rm I_{\rm C}}$ $\sim$ 0.4 (almost identical within the error bars, though), which are physically indistinguishable from those of short-period comets in the inner solar system \citep{Solontoi2012}. During the activation, however, fluxes in the $R_{\rm C}$ band became significantly enhanced, resulting in the redder color of $m_{\rm g'}$ $-$ $m_{\rm R_{\rm C}}$ and more neutral color of $m_{\rm R_{\rm C}}$ $-$ $m_{\rm I_{\rm C}}$. Such magnitude differences had remained nearly constant at the value of $m_{\rm g'}$ $-$ $m_{\rm R_{\rm C}}$ = 0.9 $\pm$ 0.1 (the standard deviation of the nominal values) and $m_{\rm R_{\rm C}}$ $-$ $m_{\rm I_{\rm C}}$ = 0.2 $\pm$ 0.1 over the following 1.5 months. After peaking the maximal $Af\rho$ and $\dot{M}_{\rm d}$, the coma color again turned to reddish-neutral, albeit with large uncertainties, consistent with the status before the activation.

\section{Orbital evolution of 252P/LINEAR and its dynamical association with P/2016 BA14 (PanSTARRS) \label{sec:orbit}}

In this section, we investigated orbital evolution of 252P and its dynamical association with P/2016 BA14 (PanSTARRS) to search an existence of any possible dynamical characteristics of the objects attributable to the observed photo-polarimetric properties of 252P. 

Comet P/2016 BA14 (PanSTARRS; hereafter ``BA14''), a possible dynamical pair of 252P with remarkably similar orbital elements \citep{Rudawska2016}, possesses a geometric albedo of <3 \% \citep{Reddy2016} and diametric size of >1 km \citep{Naidu2016} or 1--2 km \citep{Li2017}. Unlike the capricious behaviors of 252P, BA14 showed a >6 mag larger r$'$-band magnitude with a fraction of active area on the surface of $\sim$0.01 \%, i.e., being rather asteroidal in the 2016 apparition \citep{Li2017}. To trace the dynamic history of the two back to the time when they were ejected into the near-Earth orbit, we conducted a backward dynamical simulation with the Mercury 6 integrator \citep{Chambers1999} under the gravity of the Sun and eight planets. We generated 1000 clones randomly distributed within the 1-$\sigma$ Gaussian distribution, with the orbital elements quoted on the JPL Small-Body Database Browser site. They were integrated 1000 yr backwards with 8 day time step in the general Bulirsch-Stoer mode without considering the Yarkovsky force (i.e., consistent with the negligible activity levels before the 2016 apparition). The applied orbital elements and their errors are summarized in Table \ref{t4}.\\

\begin{table*}[!h]
\centering
\small
\caption{The orbital elements of 252P and BA14 and their 1-$\sigma$ uncertainties at Epoch 2457519.5 (=2016 May 11.0)}
\begin{tabular}{c|cccccc}
\toprule
Object & $a$ $^{\rm a}$ & $e$ $^{\rm b}$ & $i$ $^{\rm c}$ & $\omega$ $^{\rm d}$ & $\Omega$ $^{\rm e}$ & $M$ $^{\rm f}$ \\
\midrule
\midrule
\multirow{2}{*}{252P}  & 3.0470 & 0.6731 & 10.4223 & 343.3110 & 190.9499 & 10.5137 \\
& (6.3018 $\times$ 10$^{\rm -5}$) & (6.8756 $\times$ 10$^{\rm -6}$) & (6.3172 $\times$ 10$^{\rm -5}$) & (5.4848 $\times$ 10$^{\rm -5}$) & (6.2681 $\times$ 10$^{\rm -5}$) & (3.000 $\times$ 10$^{\rm -4}$) \\
\midrule
\multirow{2}{*}{BA14} & 3.0216 & 0.6662 & 18.9181 & 351.9022 & 180.5312 & 40.8120 \\
& (8.4014 $\times$ 10$^{\rm -7}$) & (9.2653 $\times$ 10$^{\rm -8}$) & (1.7085 $\times$ 10$^{\rm -6}$) & (4.5491 $\times$ 10$^{\rm -6}$) & (2.209 $\times$ 10$^{\rm -6}$) & (1.7161 $\times$ 10$^{\rm -5}$) \\
\bottomrule
\end{tabular}
\tablefoot{
\tablefoottext{\rm a}{ The semimajor axis in au;}
\tablefoottext{\rm b}{ The eccentricity;}
\tablefoottext{\rm c}{ The inclination in degrees;}
\tablefoottext{\rm d}{ The argument of perihelion in degrees;}
\tablefoottext{\rm e}{ The longitude of the ascending node in degrees;}
and \tablefoottext{\rm f}{ The mean anomaly in degrees.} Numbers in parentheses are 1-$\sigma$ uncertainties of the orbital elements. All the values are quoted from the JPL Small-Body Database Browser (https://ssd.jpl.nasa.gov/sbdb.cgi).}
\label{t6}
\vskip-1ex
\end{table*}

In addition, to assess the orbital similarity of two bodies in the near-Earth orbit, we estimated a traditional {\it D}$_{\rm SH}$ parameter \citep{Southworth1963}, which is a distance between the orbits of two bodies in five-dimensional orbital element space ($q$, $e$, $i$, $\omega$, and $\Omega$) calculated by
\begin{equation}
\begin{aligned}
D_{\rm SH} = {} & \Bigg[\big(e_{\rm B} - e_{\rm A}\big)^{\rm 2} + \big(q_{\rm B} - q_{\rm A}\big)^{\rm 2}  \\
& + \bigg(2 \sin \frac{I_{\rm AB}}{2} \bigg)^{\rm 2} + \bigg[\big(e_{\rm A} + e_{\rm B}\big) \sin \frac{\Pi_{\rm AB}}{2}\bigg]^{\rm 2}\Bigg]^{\rm 1/2},
\label{eq:eq15}
\end{aligned}
\end{equation}
\noindent where
\begin{equation}
I_{\rm AB} = \arccos \Big[ \cos i_{\rm A} \cos i_{\rm B} + \sin i_{\rm A} \sin i_{\rm B} \cos \big( \Omega_{\rm A} - \Omega_{\rm B} \big) \Big]
\label{eq:eq16}
\end{equation}
and
\begin{equation}
\Pi_{\rm AB} = \big(\omega_{\rm A} - \omega_{\rm B} \big) + 2 \arcsin \Bigg[\cos \frac{(i_{\rm A} + i_{\rm B})}{2} \sin \frac{(\Omega_{\rm A} - \Omega_{\rm B})}{2} \sec \frac{I_{\rm AB}}{2} \Bigg].
\label{eq:eq17}
\end{equation}

    The subscripts A and B denote the two bodies to be compared (here, A for 252P and B for BA14), $q$ is the perihelion distance in au, $e$ is the eccentricity, $i$ is the inclination, $\omega$ is the argument of perihelion, and $\Omega$ is the longitude of the ascending node. In case of $D_{\rm SH}$ $\leq$ 0.20, we tend to consider that the two bodies are probably within the association range \citep{Southworth1963}. For instance, a minimum $D_{\rm SH}$ value of the possible pair, (3200) Phaethon and 2005 UD, shows $\sim$0.04 \citep{Ohtsuka2006}. 
    
\begin{figure}[!b]
\centering
\includegraphics[width=9cm]{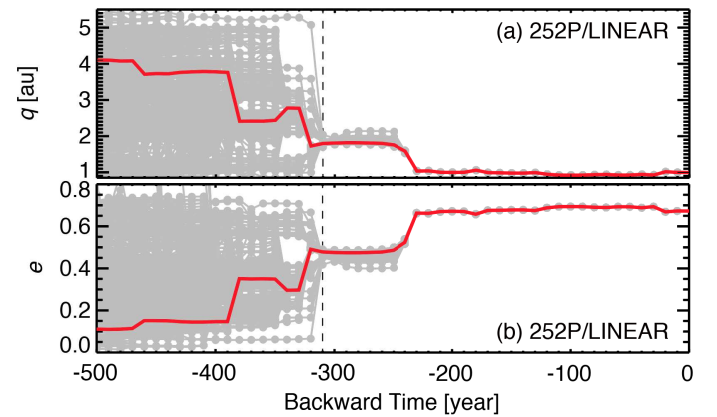}
\caption{Time evolution of (a) the perihelion distance $q$ and (b) the eccentricity $e$ of 252P. Red lines denote the evolution of a particle with the average values at the epoch (May 11.0 2016, the Julian date of 2457519.5), while each gray line represents that of an individual clone, the orbital elements of which follow 1-$\sigma$ standard deviation of the Gaussian distribution. }
\label{Fig8}
\vskip-1ex
\end{figure}

Figure \ref{Fig8} shows the time evolution of (a) the perihelion distance {\it q} and (b) the eccentricity {\it e} of 252P. Red lines denote the evolution of a particle with the average values at the epoch (May 11.0 2016, the Julian date of 2457519.5), while each gray line represents that of an individual clone, the orbital elements of which have 1-$\sigma$ standard deviation on the Gaussian distribution. Our simulation shows that 252P would have closely encountered Jupiter at the minimal distance of {\it d}$_{\rm min}$ $\sim$ 0.02 au, 314 yr ago. Due to the high uncertainty, we could only trace back the orbital evolution of the comet from the current epoch to $\sim$310 yr in the past, during which the {\it q} and {\it e} of the comet have been changed by $\delta${\it q} $\sim$ $-$1 au (decreasing) and by $\delta${\it e} $\sim$ +0.2 (increasing). Interestingly, based on the current orbital elements of 252P, the Near-Earth Object (NEO) model of \citet{Greenstreet2012} suggests two most probable source regions of the comet: the JFC source ($\sim$81.8 \%) and the outer main asteroidal belt ($\sim$16.2 \%). In either case, 252P should have eventually been transported to the current orbit by means of the resonance with Jupiter and the perturbations of terrestrial planets (e.g. \citealt{Morbidelli2002}). Based on the analysis, we conjectured that the decreasing $q$, together with its repetitive close encounters with planets (e.g., with Jupiter within $\sim$0.05 au at $\sim$320 yr ago and with the Earth within $\sim$0.02 au since $\sim$170 yr to the present), might introduce internal stresses to 252P, which would render the nucleus overall more fragile and would in turn affect the rapid polarimetric variation in its perihelion passage.

The average $D_{\rm SH}$ value derived from the osculating orbital elements of 252P and BA14 is $\sim$0.16, which implies a positive dynamical correlation of two bodies, although not as tight as that of (3200) Phaethon and 2005 UD \citep{Ohtsuka2006}. Among all the possible combinations of the 1-$\sigma$ clones, we obtained the minimal $D_{\rm SH}$ value of $\sim$0.15. Owing to the close encounters of the objects to the planets and their non-gravitational force, it is hard to know exactly when they were fragmented each other from this study.

\section{Contrast in behaviors of large and small comet nuclei in the near-Earth orbit \label{sec:size}}

This section describes our conjecture of the reason why 252P and BA14 exhibited such different activity patterns during the 2016 apparition, despite their possible dynamical association resulting in similar orbital evolution in the near-Earth orbit.

The remarkably higher activity of 252P in 2016 compared to the previous apparitions, alongside its discontinuous brightness enhancements near the perihelion, is far different from the behavior of the comet over the past apparitions. We conjecture that the effects of such orbital evolution on comets in near-Earth orbits might vary greatly depending on comet nuclei size. This conjecture would be exemplified in different observational characteristics of 252P and BA14 in the 2016 apparition, despite their similar orbital evolution over a few centuries (i.e., $D_{\rm SH}$ $\leq$ 0.20 in Appendix \ref{sec:orbit}). 252P is on the small end of sizes (300$\pm$30 m; \citealt{Li2017}) of JFCs, for which a typical comet nucleus would range from one to a few km in size \citep{Fernandez1999,Barnes2019}. The surface gravity of 252P would then be less than a factor of at least one-fifth the gravity at the surface of ordinary JFCs. The concomitant smaller escape velocity permits large particles, mm--cm in size, to readily escape the gravitational influence of 252P, while a larger nucleus would be more favorable to accumulate dust particles on its surface. 

The small size can also contribute to the spin evolution of the nuclei such that the rotational stability tends to be proportional to the fourth power of the radius of the nucleus ($\tau_{\rm ex}$ $\propto$ $r_{\rm n}^{\rm 4}$; \citealt{Jewitt2004}). 252P thus has a $\gtrsim$10$^{\rm 3}$ times smaller excitation timescale than that of a typical JFC. Indeed, a sudden increase of nongravitational force of 252P by a factor of $\sim$10 but negligible change of BA14 during the 2016 apparition \citep{Li2017} may reflect this higher rotational instability of the smaller comet. 

Taken as a whole, such consequences of the small nucleus size together with the orbital evolution in the near-Earth orbit (Appendix \ref{sec:orbit}) would enhance erratic sublimation patterns of the comet, probably resulting from highly changeable mass transfer on the surface (e.g., landslides; \citealt{Keller2017}) due to the enhanced solar radiation, and/or in unstable development of the dust mantle on the surface (e.g., \citealt{Rickman1990}). They may in turn affect the abnormal variations of the photo-polarimetric parameters of 252P near its perihelion.

\end{appendix}

\end{document}